\documentclass[11pt,,a4paper]{article}
\usepackage{amsmath}
\usepackage{amsfonts}
\usepackage{amssymb}
\usepackage{color}
\usepackage{graphicx}
\usepackage{subcaption}
\usepackage[all]{xy}
\usepackage{color}
\usepackage{authblk}
\usepackage[T1]{fontenc}
\usepackage{lmodern}
\usepackage[utf8]{inputenc}
\usepackage{graphicx}

\title{Quantum cosmology with many fluids and the choice of cosmological time}

\author[1]{G. A. Monerat\thanks{e-mail: \texttt{germano.monerat@pq.cnpq.br}}}

\author[1]{C. G. M. Santos\thanks{e-mail: \texttt{cassia.gmsantos@gmail.com}}}

\author[2]{F. G. Alvarenga\thanks{e-mail: \texttt{flavio.alvarenga@ufes.br}}}

\author[2]{S. V. B. Gon\c{c}alves\thanks{e-mail: \texttt{sergio.v.goncalves@ufes.br}}}

\author[2]{R. Fracalossi\thanks{e-mail: \texttt{rfracalossi@gmail.com}}}

\author[3]{E. V. Corr{\^e}a Silva\thanks{e-mail: \texttt{eduardo.vasquez@pq.cnpq.br}}}

\author[4]{G. Oliveira-Neto\thanks{e-mail: \texttt{gilneto@fisica.ufjf.br}}}

\affil[1]{Departamento de Modelagem Computacional, Universidade do Estado do Rio de Janeiro, Instituto Polit\'ecnico do Rio de Janeiro,  CEP 28.625-570, Nova Friburgo - RJ - Brazil.}

\affil[2]{Departamento de F\' \i sica, Centro de Ci\^encias Exatas, Universidade Federal do Esp\' \i rito Santo, CEP 29075-910, Vit\'oria, ES, Brazil.}

\affil[3]{Departamento de Matem\'atica, F\' \i sica e Computa\c c\~ao, Universidade do Estado do Rio de Janeiro, CEP 27537-000, Resende, RJ, Brazil.}

\affil[4]{Departamento de  F\' \i sica,  Instituto  de  Ci\^encias Exatas, Universidade Federal de Juiz de Fora, CEP 36036-330 - Juiz de Fora, MG, Brazil.}

\date{}
\begin{document}
\maketitle

\begin{abstract}
In this work we propose the quantization of a cosmological model describing the primordial universe filled with five barotropic fluids, namely: radiation, dust, vacuum, cosmic strings and domain walls. We intend to identify which fluid is best suited to provide phenomenologically the temporal variable in accordance with the observable universe. Through the Galerkin spectral method and the finite difference method in the Crank-Nicolson scheme (vacuum case), the quantum cosmological solutions are obtained and compared. We, also, compare the quantum cosmological solutions with the corresponding classical ones. The vacuum case is especially interesting because it provides a tunneling transition mechanism from the quantum to the classical phase and the possibility of calculating
quantum tunneling probabilities.
\end{abstract}

PACS number(s): 98.80.-k, 98.80.Cq, 04.30.-w \vspace{0.7cm}

\section{Introduction}

The fact that there is no complete theory of quantum gravity implies the need to test the quantum effects in different regimes and models of the universe. In this sense, quantum cosmological models \cite{Halliwell} are simple examples in which ideas of quantum gravitational phenomena can be tested. Using the approach known as canonical quantization, the scenario obtained is a minisuperspace  in which an infinite number of degrees of freedom are frozen, and the remainders quantized. Canonical quantization consists of describing the dynamics of the universe through the evolution of its geometric degrees of freedom and the fields present at each instant through the foliation of four-dimensional space-time in successive three-dimensional manifolds, each associated with an instant of time, a procedure performed through the so-called ADM formalism \cite{ADM}. However, the foliation procedure carries with it a negative consequence: a temporal variable becomes absent in theory. This fact is known as the problem of time in quantum cosmology \cite{Anderson}.

The nonexplicit presence of a variable of the temporal type can be overcome by the phenomenological introduction of dynamic variables associated with different material contents of the universe. The pressure of this fluid is expressed in terms of velocity potentials, of which one of them can exert the function of time. This is the so-called Schutz's formalism \cite{Schutz, Furtado}. Thus, a time-dependent
Schr\"odinger equation is established, which makes it possible to obtain the so-called universe wave function and expected values of the scale factor associated with this universe model.

In cosmology, the equation of state is the name of the equation expressing the relationship between pressure $p$ and energy density $\rho$. It is a particular example of the equation of state of any statistical mechanical system, which usually involves other variables like temperature. But in cosmology one makes the simplifying assumption that the energy density and the pressure are simply related, and that the temperature does not appear in the equation, nor any other thermodynamic variable. This is for the sake of simplicity and also because it covers a lot of interesting cases. So it is simply $p = \omega\rho$ in which $\omega$ is a number. The $\omega$ parameter could change with time, but we will assume that any time derivatives of $\omega$ are negligible compared to time derivatives of others variables. This is reasonable if the equation of state is related by microphysical processes that is not directly determined to the expansion of the universe. Thus, in a simple and efficient way for the purposes of this work the equation of state expresses that the pressure $p$ is equal to $\omega$ times the energy density $\rho$.

The situation in which the behaviour of a quantum universe whose material content is composed of a barotropic fluid has already been widely explored in the scientific literature \cite{Julio, monerat11, BoseE}. We also find cosmological models in which the material content of the universe is composed of two barotropic fluids \cite{Doisfluidos, Tovar}. In  \cite{Doisfluidos}, a universe filled with a fluid of stiff matter and radiation is quantized. In \cite{Tovar}, the quantization of a model of radiation and dust was carried out following the causal interpretation of quantum mechanics. In both cases, only the situation in which the radiation fluid is associated with time in theory is studied, and thus, non-singular universes are obtained. In this way, this article presents a significant extension of the aforementioned works, since here cases are studied in which the role of time is played by other fluids (and not only by radiation). Increasing the number of degrees of freedom of the models enriches the description of the different cosmological scenarios and may lead to greater accuracy. As in the Wheeler-DeWitt equation the absence of the time parameter is a very important theoretical problem, all the cited models attempt to circumvent this using the most varied methods. The Schutz's formalism  is very useful for solving this fundamental question. In this formalism the material content plays the role of the temporal variable. But that choice is quite arbitrary. 

Here in this work we intend to analyze the behavior of the quantum universe where the material content is composed of five barotropic fluids. Our objective is to study the scale factor behaviour and the states of the primordial universe in a model that, in relation to the more conventional works, has a greater number of degrees of freedom associated with matter. Furthermore, we hope to check which of the fluids used as the temporal variable produces the best behavior for the universe in the sense of its passage to the classical phase of its evolution.

The structure and organization of the article is elaborated as follows: in Sec. $2$ we introduce the characteristics of the model, explaining the role of each barotropic fluid in the evolution of the universe, namely: cosmic strings, domain walls, radiation and dust; in Sec. $3$  the vacuum case is analyzed separately. We compute the expected values and classical evolution for the scale factor of the universe. Finally, in Sec. $4$ we present our conclusions and discuss our results.

\section{The Model}
\label{CM}

In this model we consider a homogeneous and isotropic Friedmann-Lemaitre-Robertson-Walker (FLRW) universe with positively curved spatial sections:

\begin{equation}
ds^2=-N(t)^2dt^2+a(t)\left(\frac{dr^2}{1-r^2}+r^2d\theta^2+r^2sin^2\theta d\phi^2\right)\;,\label{FRW}
\end{equation}
 \noindent in which $a(t)$ is the scale factor and $N(t)$ denotes the lapse function. Here, we are using the natural
unit system, where $\hbar=c=16\pi G=1$.

We start with the Einstein-Hilbert action with a boundary term plus matter
\begin{align}
S =\int_M d^{4}x\sqrt{-g}R+2\int_{\partial M} d^{3}x\sqrt{h}K + \int\,d^4x\sqrt{-g}\,p\quad,\label{fk1}
\end{align}
\noindent in which $R$ is the scalar curvature,  $K$ is the trace of the the extrinsic curvature $K_{ab}$, $h$  is the determinant of the induced metric $h_{ab}$ over the three-dimensional spatial hypersurface, which is the boundary $\partial M$ of the four dimensional manifold $M$ and $p$ is the pressure. 


The last term in (\ref{fk1}) is Schutz's action \cite{Schutz} that describes the dynamics of a relativistic perfect fluid in interaction with the gravitational field in terms of velocity potentials. It assumes that in the absence of rotation, the four-velocity can be expressed in terms of potentials $\phi, \theta$ and $S$ as 

\begin{equation}
U_{\nu} = \frac{1}{\mu}(\phi_{,\nu} +  \theta S_{,\nu}) \label{potencial}\, .
\end{equation}

\noindent The variable $\mu$ is the specific enthalpy,  $S$ is the specific entropy, whereas the potentials $\phi$ and $\theta$ have no clear physical meaning. The four-velocity obeys the normalization condition

\begin{equation}
U^{\nu}U_{\nu}=-1\, \, ,\label{131}\,\,
\end{equation}

\noindent so that we can express $\mu$ in terms of the potentials

\begin{equation}
\mu=\frac{1}{N}(\dot{\phi}+\theta\dot{S})\,\,\label{entalpia2} .
\end{equation}

The basic thermodynamic relations for a barotropic perfect fluid, $p=\omega\rho$, are given by

\begin{equation}
\label{termo}
\rho=\rho_0(1+\Pi)\, ;\, \quad \mu=(1+\Pi) + \frac{p}{\rho_{0}} \, ; \, \quad \Theta dS=d\Pi +p d\bigg(\frac{1}{\rho_{0}}\bigg) \, .
\end{equation}

\noindent Here $\Pi$ is the specific internal energy and $\Theta$ is the temperature. By the identity

\begin{equation}
d\Pi +p d\bigg(\frac{1}{\rho_{0}}\bigg)=(1+\Pi)d[\hbox{ln}(1+\Pi)-\omega\,\hbox{ln}\rho_0]\,\, ,
\end{equation}

\noindent we can identify 

\begin{equation}
\label{entropia}
\Theta=1+\Pi\,\,\, ; \,\,\, S=\hbox{ln}\frac{(1+\Pi)}{{\rho_{0}}^{\omega}}\,\, . 
\end{equation}

If we now combine the equations (\ref{termo}) and (\ref{entropia}),  we can by means of the equation of state express the pressure as

\begin{equation}
 p=\omega\left(\frac{\mu}{\omega+1}\right)^{1+\frac{1}{\omega}}e^{-\frac{S}{\omega}}\,\, \label{pressure}.
\end{equation}

By using (\ref{FRW}) and (\ref{pressure}) in the action (\ref{fk1}), it is possible to obtain the Lagrangian

\begin{equation}
L =  - 6\frac{\dot{a}^2\,a}{N} + 6Na + N^{-\frac{1}{\omega}}\,\,a^3\,\frac{\omega}{(\omega+1)^{1+\frac{1}{\omega}}}\,\,(\dot{\phi}+\theta\dot{S})^{1+\frac{1}{\omega}}\,\, e^{-\frac{S}{\omega}}\,\, .
\end{equation}

The conjugate momenta are derived from the above Lagrangian, written in terms of the canonical variables
\begin{eqnarray}
p_a&=&-12 \frac{a\dot{a}}{N}\quad,\nonumber\\
p_{\phi}&=& N^{-1/{\omega}}\,\, a^3\,\, \frac{1}{({\omega}+1)^{1/{\omega}}} \,\,(\dot{\phi}+
{\theta}\dot{S})^{1/{\omega}}\,\,  e^{-S/{\omega}}\quad,\nonumber\\
p_S&=&\theta p_{\phi}\quad,\nonumber\\
p_{\theta}&=&0\quad .
\end{eqnarray}

Finally, if we use the canonical formalism the action (\ref{fk1}) reduces to

\begin{equation}
{\cal S}=\int_{}^{}dt\{\dot{a}p_a + \dot{S}p_S - N\cal{H}\}\quad ,
\label{geral01}
\end{equation}

\noindent in which the super-Hamiltonian $\cal{H}$ is

\begin{equation}
{\cal{H}}=-\frac{p_a^2}{24a}-6a + {p_{\phi}}^{\omega+1}\,\,\frac{e^S}{a^{3\omega}}\,\,\,.\label{hS}
\end{equation}

The canonical transformation
\begin{displaymath}
T=p_S\, e^{-S}\,{p_{\phi}}^{-(\omega+1)}\,\,\quad ; \, \, \quad p_{T} = {p_{\phi}}^{(\omega + 1)}\, e^{S}\, ; \, 
\end{displaymath}
\begin{equation}
 \overline{\phi}=\phi -(\omega+1)\frac{p_{S}}{p_{\phi}}\, \,\quad; \, \,\quad
\overline{p}_{\phi}=p_{\phi}\,\,\quad , \label{tc1}
\end{equation}
\noindent allows us to introduce the moment associated with the fluid variable varying linearly in (\ref{hS}). With the choice of the lapse function given by $N = a$, the Hamiltonian of a single perfect fluid can be written as follows
\begin{equation}
H=N{\cal H}=-\frac{p_{a}^{2}}{24}-6 a^2 + {p_{T}}{a^{1-3\omega}}\quad.
\label{geral02}
\end{equation}

For this  Hamiltonian to be self-adjoint, one has to introduce a weight function
in the inner product of two wave functions $\Phi$ and $\Psi$, which reads

\begin{equation}
\left(\Phi,\Psi\right)=\int_{0}^{\infty}a^{1-3\omega}\;\Phi^{\star}\Psi da.
\label{produto}
\end{equation}

According to the type of fluid, the parameter $\omega$ in its equation of state takes specific values\footnote{Here and in what follows, the subscripts $r$, $d$, $cs$, $dw$ and $v$ represent, respectively: radiation, dust, cosmic strings, domain walls and vacuum.}:
$\omega_{r}=1/3$ for radiation; 
$\omega_{d}=0$ for dust; 
$\omega_{cs}=-1/3$ for cosmic strings; 
$\omega_{dw}=-2/3$ for domain walls; and
$\omega_{v}=-1$ for vacuum.

The Hamiltonian of a cosmological model with all those five fluids can be extended from (\ref{geral02}) to

\begin{equation}
H = - \frac{{p_{a}}^{2}}{24}-6 a^2 + p_{T_{r}} + p_{T_{d}}\,a + p_{T_{cs}}\,a^2 + p_{T_{dw}}\,a^3 +  p_{T_{v}}\,a^4\quad.
\label{cincofluidos}
\end{equation}

We can now implement the canonical quantization procedure in minisuperspace and obtain the corresponding Wheeler-DeWitt equation
\begin{equation}
\label{WDW01}
{\hat H}\Psi=0\quad.
\end{equation}

Substituting the momenta by their corresponding operators
\begin{equation}
\label{momentaoperator}
{\hat{p}}_a \rightarrow - i \frac{\partial}{\partial a}\,\,\quad \hbox{and} \,\,\quad {\hat{p}}_{T_j} \rightarrow - i \frac{\partial}{\partial T_j}\quad ,
\end{equation}
\noindent we can transform the Wheeler-DeWitt equation (\ref{WDW01}) into a genuine time dependent
Schrödinger equation as a consequence of the linear contribution of the momenta associated with the variable $T$. In eq. (\ref{momentaoperator}), the index $j$ varies from 0 until 4, one for each fluid:
$j=0$ (radiation), $j=1$ (dust), $j=2$ (cosmic strings), $j=3$ (domain walls), $j=4$ (vacuum).
Therefore, we are going to quantize all five momenta, obtaining a Schrödinger equation with
five time variables, one for each fluid.
\begin{equation}
\label{fivetimesschrodinger}
- \frac{\partial^2 \Psi(a,t_j)}{\partial a^2} + 144a^2 \Psi(a,t_j) = 24 i \sum_{j=0}^{j=4} a^j \frac{\partial \Psi(a,t_j)}{\partial t_j}\quad,
\end{equation}
In Eq. (\ref{fivetimesschrodinger}) the time parameters have been rescaled as $(t_j\rightarrow - T_j)$.
Then, we shall write the wavefunction $\Psi(a,t_j)$, as a product of six different functions: one for each time variable plus one for the scale factor,
\begin{equation}
\label{separavariaveis}
\Psi(a,t_j) = \Phi (a) \prod_{j=0}^{j=4} e^{-i E_j t_j}.
\end{equation}
Next, we introduce that wavefunction in equation (\ref{fivetimesschrodinger}), with five time variables and choose one of the time variables. For all other time variables, we compute the corresponding time derivatives of the wavefunction and obtain
the following Schrödinger equation, 
\begin{equation}
\label{onetimeschrodinger}
- \frac{\partial^2 \Psi(a,t_m)}{\partial a^2} + V_{\rm efm}(a) \Psi(a,t_m) = 24 i a^m \frac{\partial \Psi(a,t_m)}{\partial t_j}\quad,
\end{equation}
where $m$ is one of the possible values of $j$ and $V_{\rm efm}(a)$ is an effective potential that depends on the time variable chosen. In this way,
it is possible to introduce time in five different ways, each one associated with a specific fluid. Next, in the present section, we will present and investigate four possible time variable choices, one for each perfect fluids: cosmic strings, domain walls, radiation and dust. The vacuum case will be analyzed in the following section.

\subsection{Cosmic strings fluid $(\omega_{cs} = - 1/3)$}

If  we have an exotic matter with an equation of state of the type $p_{cs}  = -\frac{1}{3}\,\rho_{cs}$, this fluid is called cosmic strings fluid. They were first introduced by theoretical physicist Tom W. B. Kibble \cite{Kib01, Kib02} in the late $70$s as a possible result of some field theories. They are objects that may have formed in the early Universe, through a phase transition, with a one-dimensional (line-like) structure. Cosmic strings were a popular research topic in the $80$s, since they could have triggered the formation of large-scale structures such as galaxies \cite{Vil}. The  presence of cosmic strings in the early universe would leave an imprint in the cosmic microwave background (CMB). Space-based experiments like COBE and WMAP revealed that cosmic strings do not make a measurable contribution to the CMB, thus ruling out a significant role for cosmic strings \cite{Smoot}.

This kind of matter was revisited in the early $2000$s when it was realized that it could also be formed in the context of string theory \cite{Sarangi,Copeland,Sakellariadou}, in
which elementary particles are described by tiny one-dimensional objects in a multi-dimensional space. In some theories, the strings could grow to cosmological scales and behave like historical cosmic strings. These are called cosmic superstrings, and could provide precious observational signatures of string theory. Assuming they do exist, the network of cosmic (super)strings formed in the early universe would have evolved as the universe expanded. Cosmic strings, though ruled out as the origin of cosmological structure, have recently obtained renewed popularity by the recognition that a variety of string theory-motivated and hybrid models for inflation generically predict the formation of cosmic string networks \cite{Pillado}. Strings are limited to producing less than about $10\%$ of the primordial CMB anisotropy. CMB data can actually favor a contribution from strings if the inflationary spectrum is exactly Harrison-Zeldovich ($n_s = 1$) \cite{Moss}. Thus, the existence of cosmic strings remains a serious element in the context of cosmology, mainly in the early phase of the evolution of the universe.
Here, in this work, the cosmic string networks is considered as one of the basic elements refering to the material content of the universe. 

Now, if we choose the cosmic strings time, eq. (\ref{onetimeschrodinger}) may be written as,
\begin{equation}
\label{ewdcs}
- \frac{\partial^2 \Psi(a,t)}{\partial a^2} + V_{\rm efcs}(a) \Psi(a,t) = 24 i \, a^2 \frac{\partial \Psi(a,t)}{\partial t}\quad,
\end{equation}
for the wave function of the universe, in which $V_{efcs}(a)$ is the effective potential given by,
\begin{equation}
V_{\rm efcs}(a)= 144 a^2 -24 E_{r} -24 E_{d}\,a -24 E_{dw}\,a^3 -24 E_{v}\,a^4\quad.
\label{pecordas}
\end{equation}

\begin{figure}[h!] 
\begin{subfigure}{0.5\textwidth}
\includegraphics[width=0.9\linewidth, height=5cm]{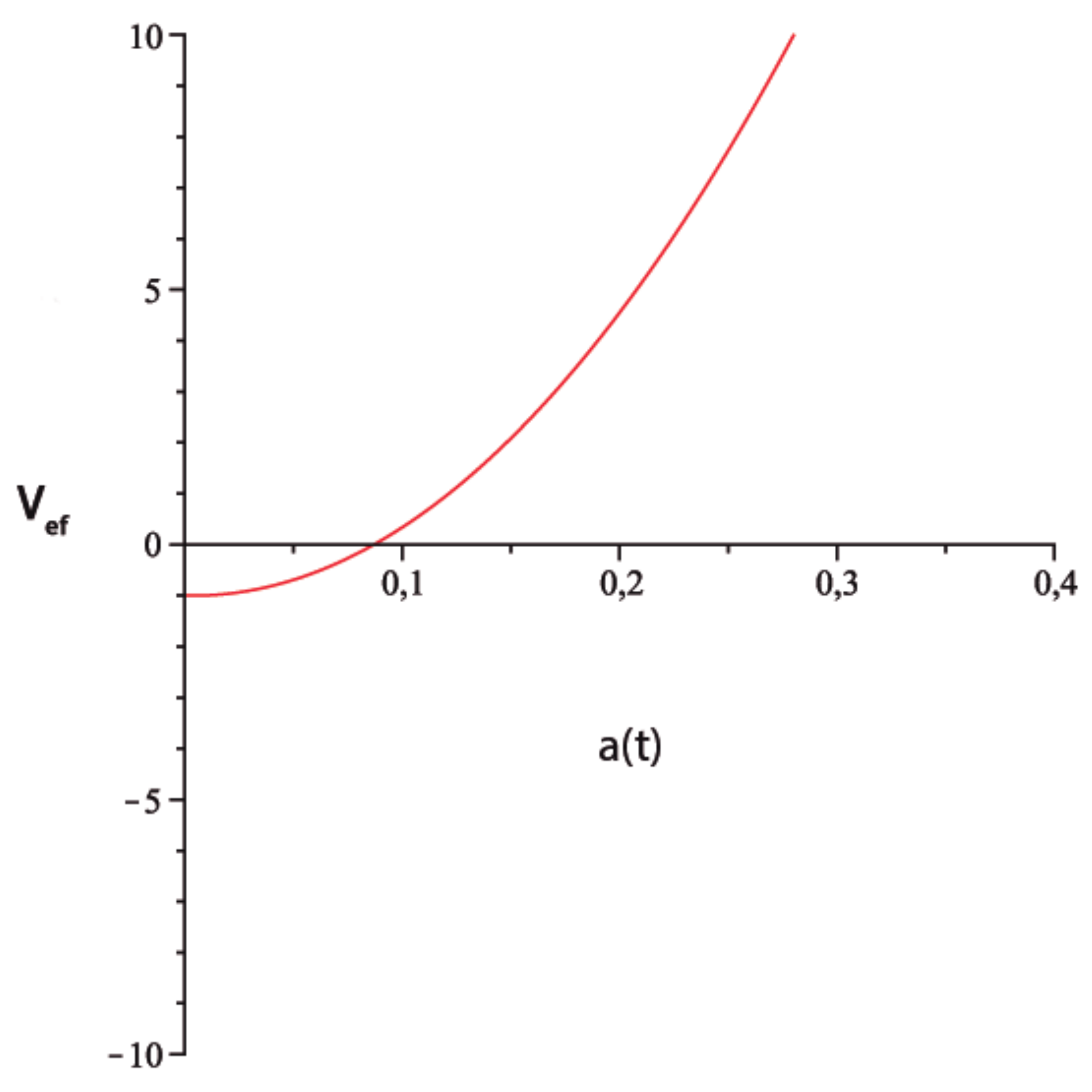} 
\caption{Quantum potential for the cosmic string model.}
\end{subfigure} 
\begin{subfigure}{0.5\textwidth}
\includegraphics[width=0.9\linewidth, height=5cm]{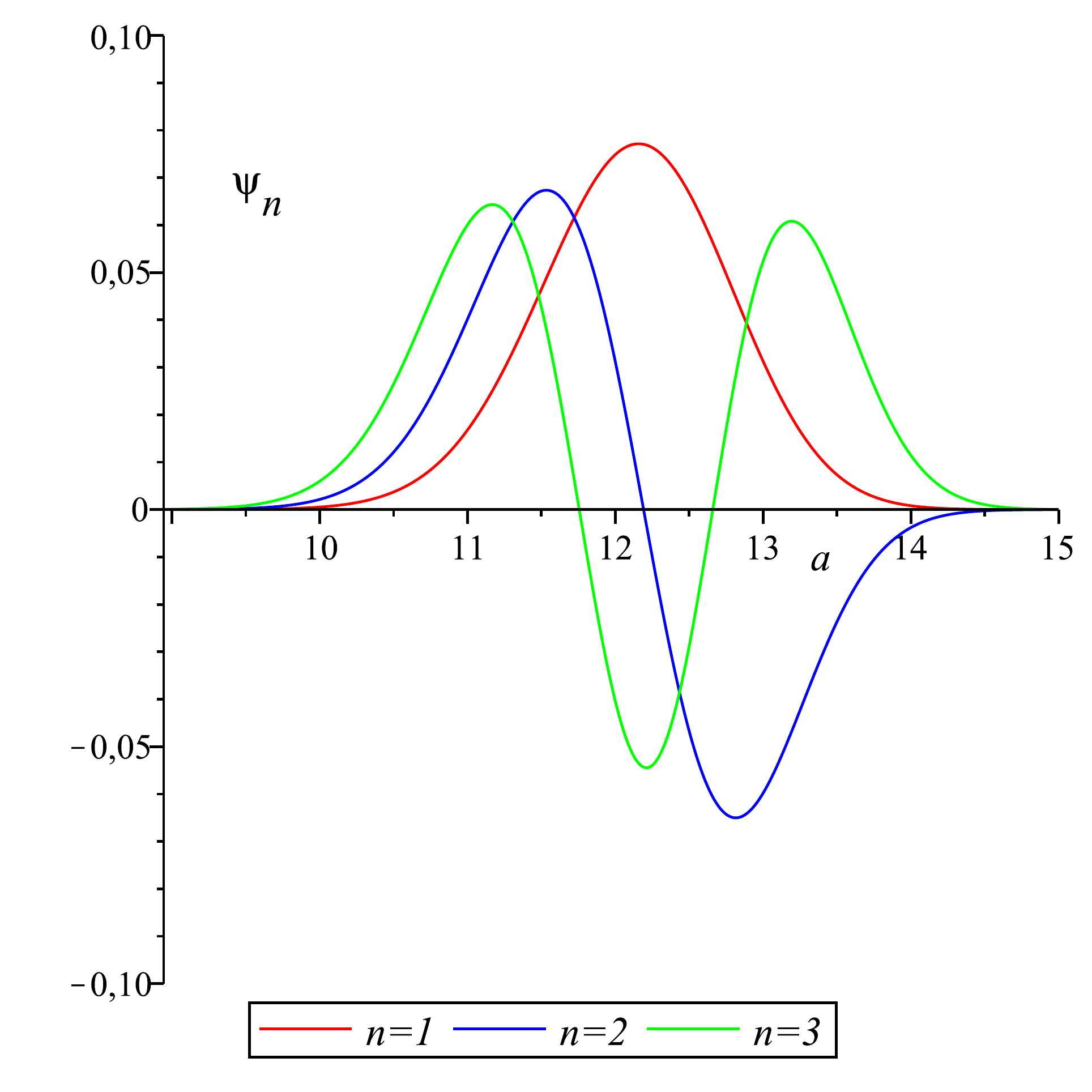}
\caption{Examples of eigenstates for the cosmic strings fluid.}
\end{subfigure}
\begin{subfigure}{0.5\textwidth}
\includegraphics[width=0.9\linewidth, height=5cm]{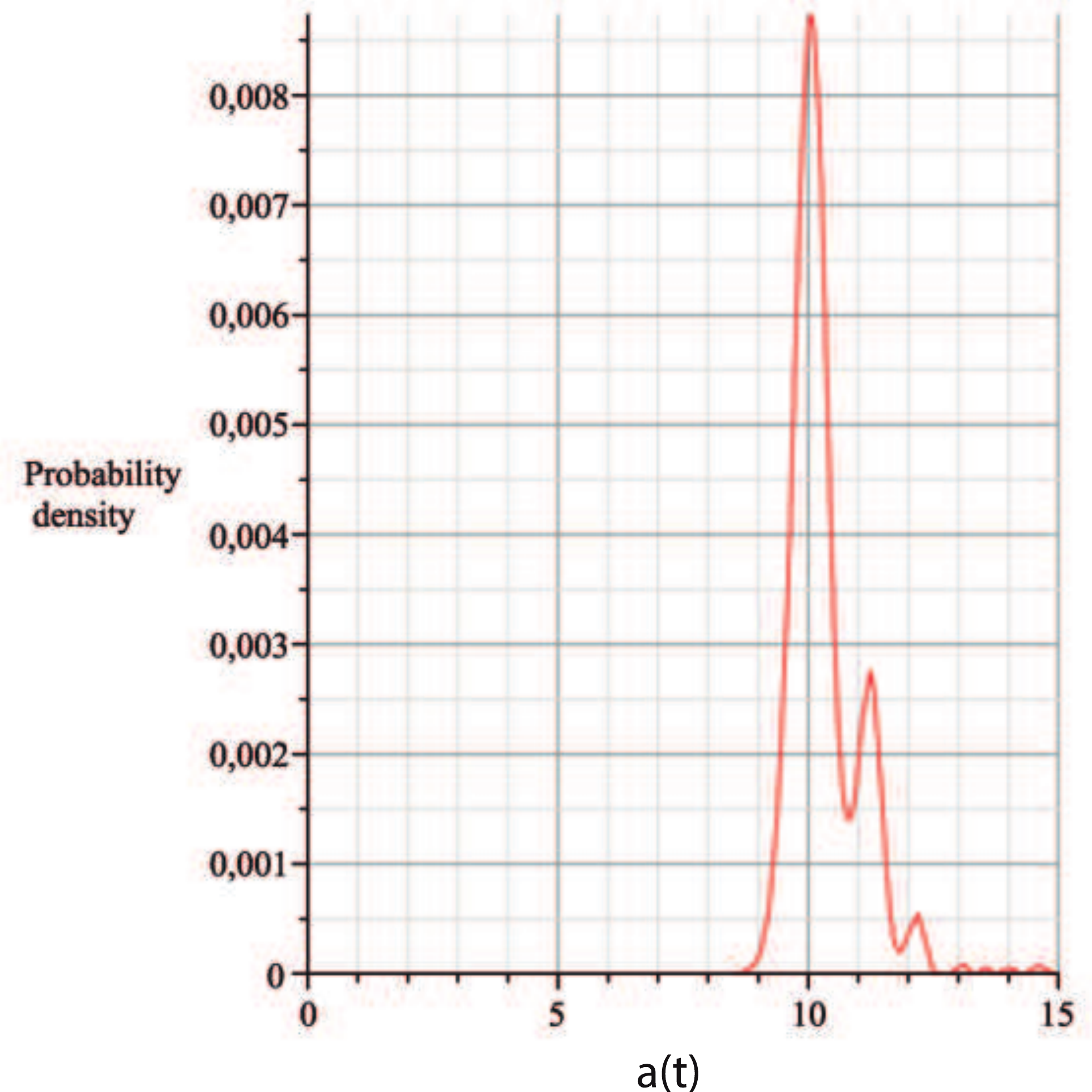} 
\caption{Initial probability density  $|\Psi(a,0)|^2$}
\end{subfigure}
\begin{subfigure}{0.5\textwidth}
\includegraphics[width=0.9\linewidth, height=5cm]{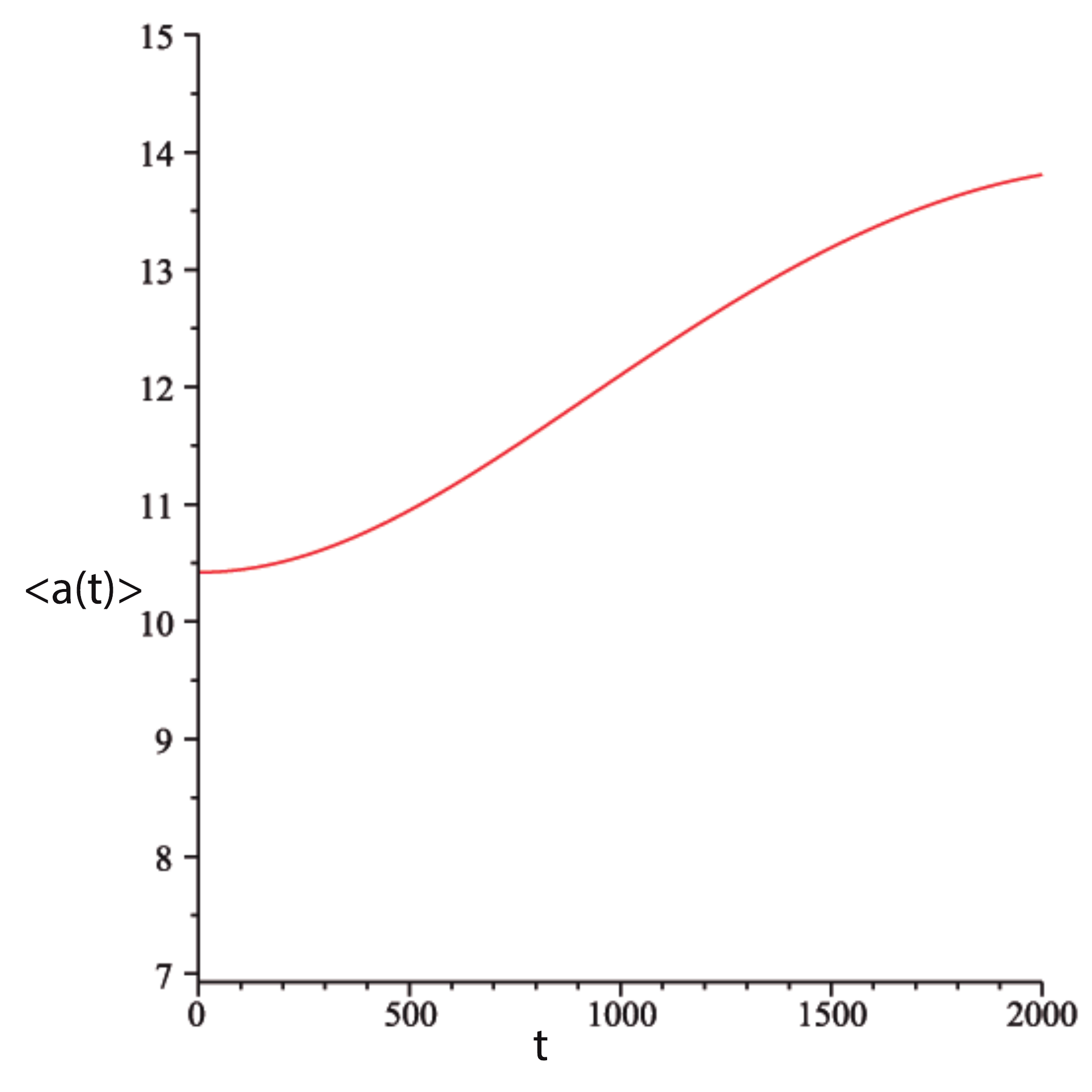} 
\caption{Expected value $\left<a\right>$, as a function of time $t$ for the cosmic strings fluid.}
\end{subfigure}
\caption{Cosmological solutions for the case of a cosmic string fluid.}
\label{csfig}
\end{figure}

\begin{figure}[h!]
\includegraphics[width=0.9\linewidth, height=5cm]{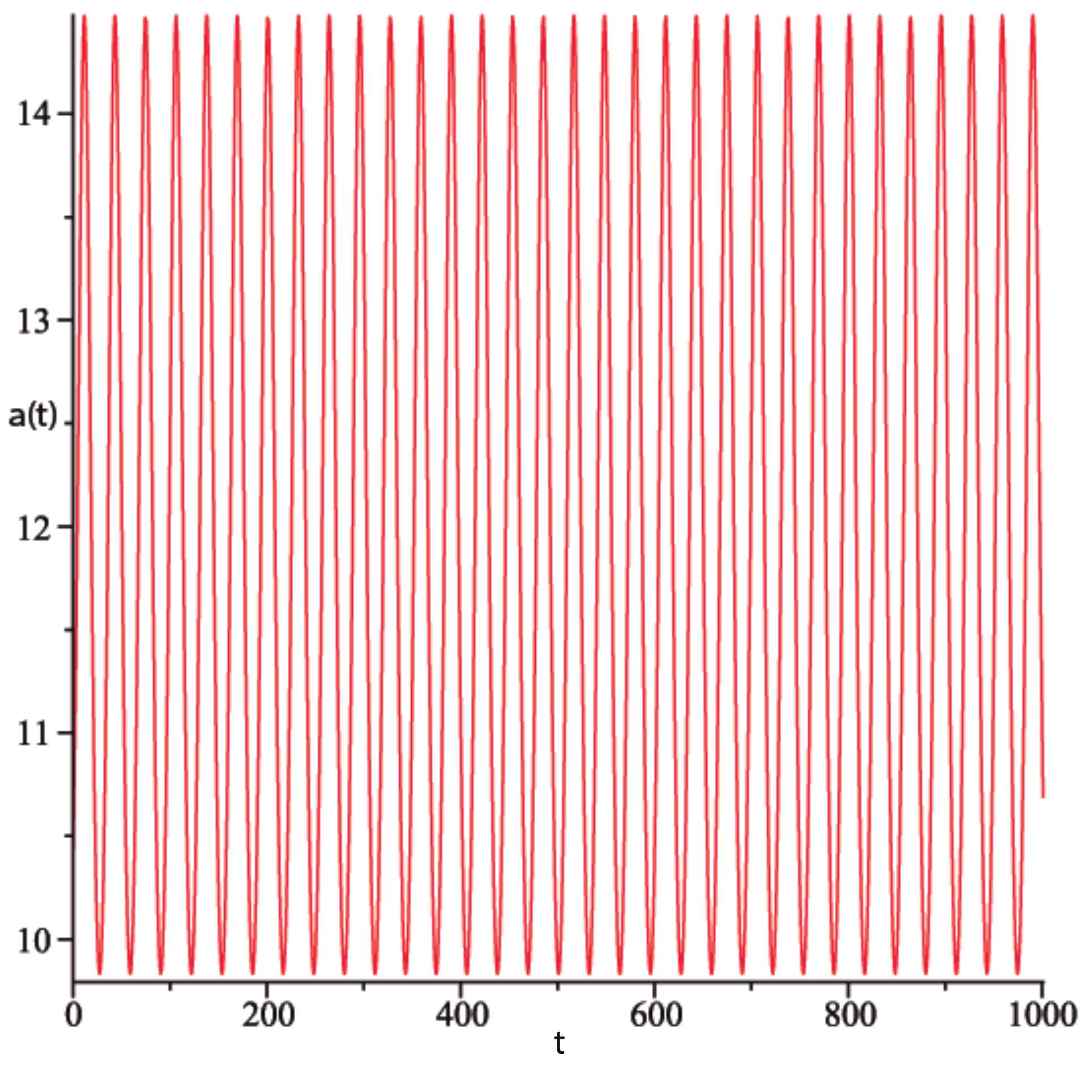} 
\caption{classical scale factor behavior for the corresponding
model to the one shown in Figure \ref{csfig}}
\label{classicalcs}
\end{figure}

Notice that (\ref{ewdcs}) is not a Schr\"{o}dinger equation, because
the term that involves a time derivative has a factor $a^2$. Since, the 
energies $E_j$ of the others fluids, in the potential $V_{efcs}(a)$ eq. (\ref{pecordas}),
were not quantized, they have a continuum spectrum. The system
has bound states and its approximate solution has been obtained in
\cite{Monerat} using the spectral method \cite{Boyd}. A finite interval
$[0,L]$, with $L>0$, has been used; once obtained the energy spectrum
and its eigenfunctions, wave packages of finite norm has been obtained
by the superposition of the 10 lower-level states. Each eigenfunction,
as well as the wave packages obtained, vanish at $a=0$ e $a=L$. Figure
\ref{csfig} shows an example for the present case. There, we can see 
the expected value of the scale factor of the Universe, the first three 
eigenstates, as well as the initial probability density and the effective 
potential obtained. In this example we used: $E_{v}=-0.0017,\,\, E_{d}=1/24,\,\, E_{dw}=1/24,\,\, E_{r}=1/24,\,\, L=15$.
In Figure \ref{classicalcs}, we show the classical scale factor behavior for the corresponding
model to the one shown in Figure \ref{csfig}. In the classical model we used:
$p_{T_{v}}=-0.0017,\,\, p_{T_{d}}=1/24,\,\, p_{T_{dw}}=1/24,\,\, p_{T_{r}}=1/24$, which are identical to the corresponding energies in the quantum model.
For $p_{T_{cs}}$, we considered the mean value of the ten eigenvalues used
in order to construct the wavepacket of the quantum model: $p_{T_{cs}}=5.75$. 
As the scale factor initial value ($a(0)$), we used the scale factor expected value at $t=0$: $a(0)=10.4206808971500$.
Finally, the initial value for the scale factor time derivative ($\dot{a}(0)$) was obtained by using the appropriate Friedmann equation
with the values given above for $p_{T_{v}}, p_{T_{d}}, p_{T_{dw}}, p_{T_{r}}, p_{T_{cs}}, a(0)$: $\dot{a}(0)=0.2681578789$.
Comparing Figures \ref{csfig}(d) and \ref{classicalcs}, we notice that the scale factor expected value is oscillating, such that, its maxima and minima values are inside the region where the classical scale factor trajectory oscillates.


\subsection{Domain walls fluid $(\omega_{dw} = -2/3)$}

As in the case of the cosmic strings, the presence of a particular defect after spontaneous symmetry breaking is determined by the topology of the vacuum manifold of the theory or model in question. Strings are line-like defects which form if an axial or cilindrical symmetry is broken. Among those defects, domain walls are the simplest since they arise if the potential of a field has a global discrete symmetry that is spontaneously broken by the vacuum. It occurs when the vacuum manifold has two or more disconnected components \cite{Rubakov}. In our cosmological context, the universe after the phase transition divides into domains, each populated at random by one of the available vacua. In recent times the motivation for the study of cosmic defects in general and domain walls in particular arise in the context of branes theories \cite{Daniele}.

In terms of hydrodynamical description the domain walls fluid is represented by the particular equation of state $p_{dw} = - \frac{2}{3}\,\rho_{dw}$. This type of fluid is also known as solid dark energy, which possesses resistance to pure shear deformations, guaranteeing stability with respect to small perturbations \cite{SergioJ}. The microphysical origin for solid dark energy is a dense network of low-tension domain walls. This is attractive for several reasons. First, domain walls are ubiquitous in field theory, inevitably appearing in models with spontaneously broken discrete symmetries. Second, domain walls, and the solid dark energy in general, have been shown to be compatible with the observations of large scale structure. Finally, a static wall network has an equation of state $\omega_{dw} = - 2/3$, consistent with all observational data.

Now, if we choose the domain walls time, eq. (\ref{onetimeschrodinger}) may be written as,
\begin{equation}
\label{ewddw}
- \frac{\partial^2 \Psi(a,t)}{\partial a^2} + V_{\rm efdw}(a) \Psi(a,t) = 24 i\, a^3\, \frac{\partial \Psi(a,t)}{\partial t}\, ,
\end{equation}
in which the effective potential is given by
\begin{equation}
V_{\rm efdw}(a)= 144 a^2 -24 E_{r} -24 E_{d}\,a -24 E_{cs}\,a^2 -24 E_{v}\,a^4 \,.
\label{peparedes}
\end{equation}

\begin{figure}[h!]
\begin{subfigure}{0.5\textwidth}
\includegraphics[width=0.9\linewidth, height=5cm]{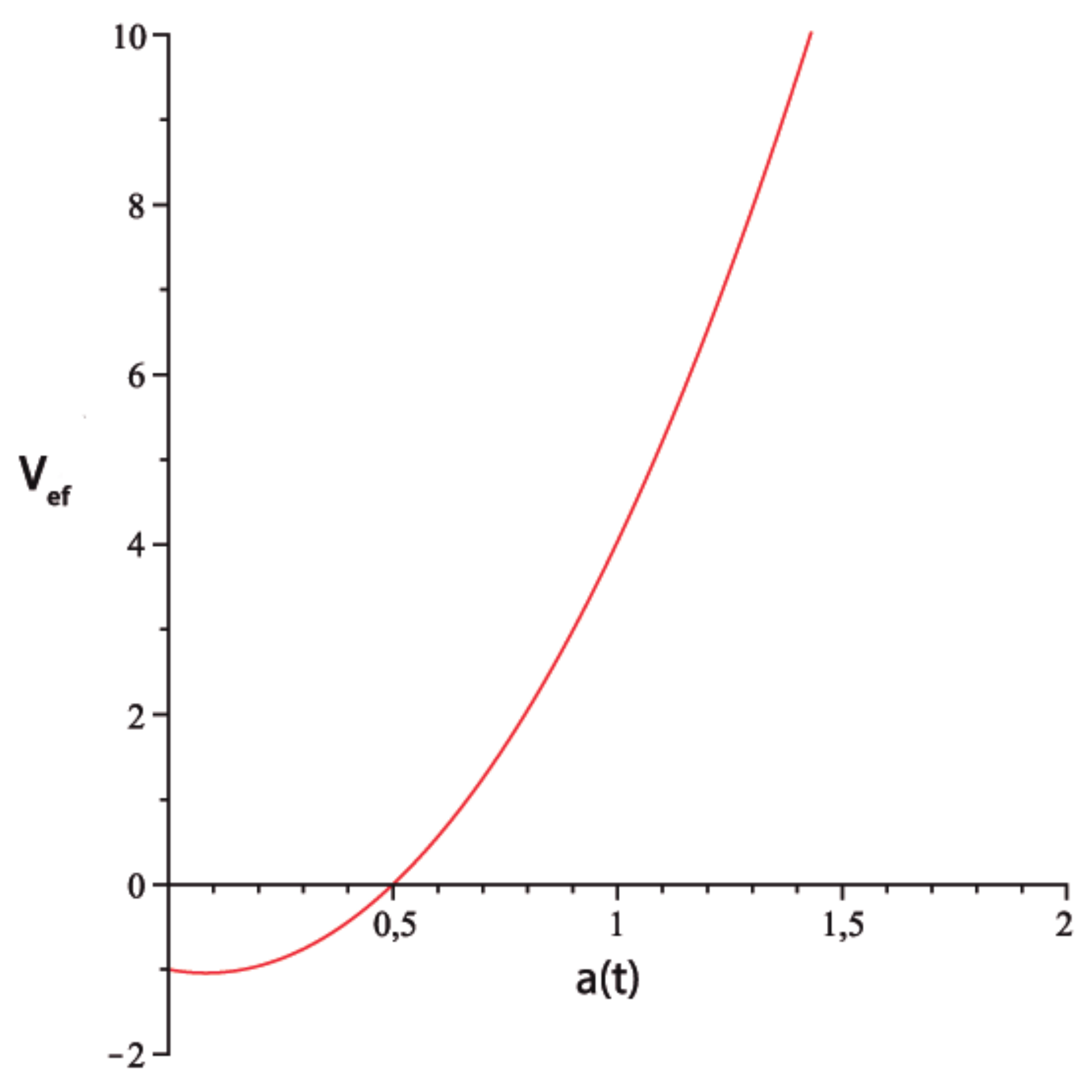} 
\caption{Effective potential for the domain walls model.}
\end{subfigure}
\begin{subfigure}{0.5\textwidth}
\includegraphics[width=0.9\linewidth, height=5cm]{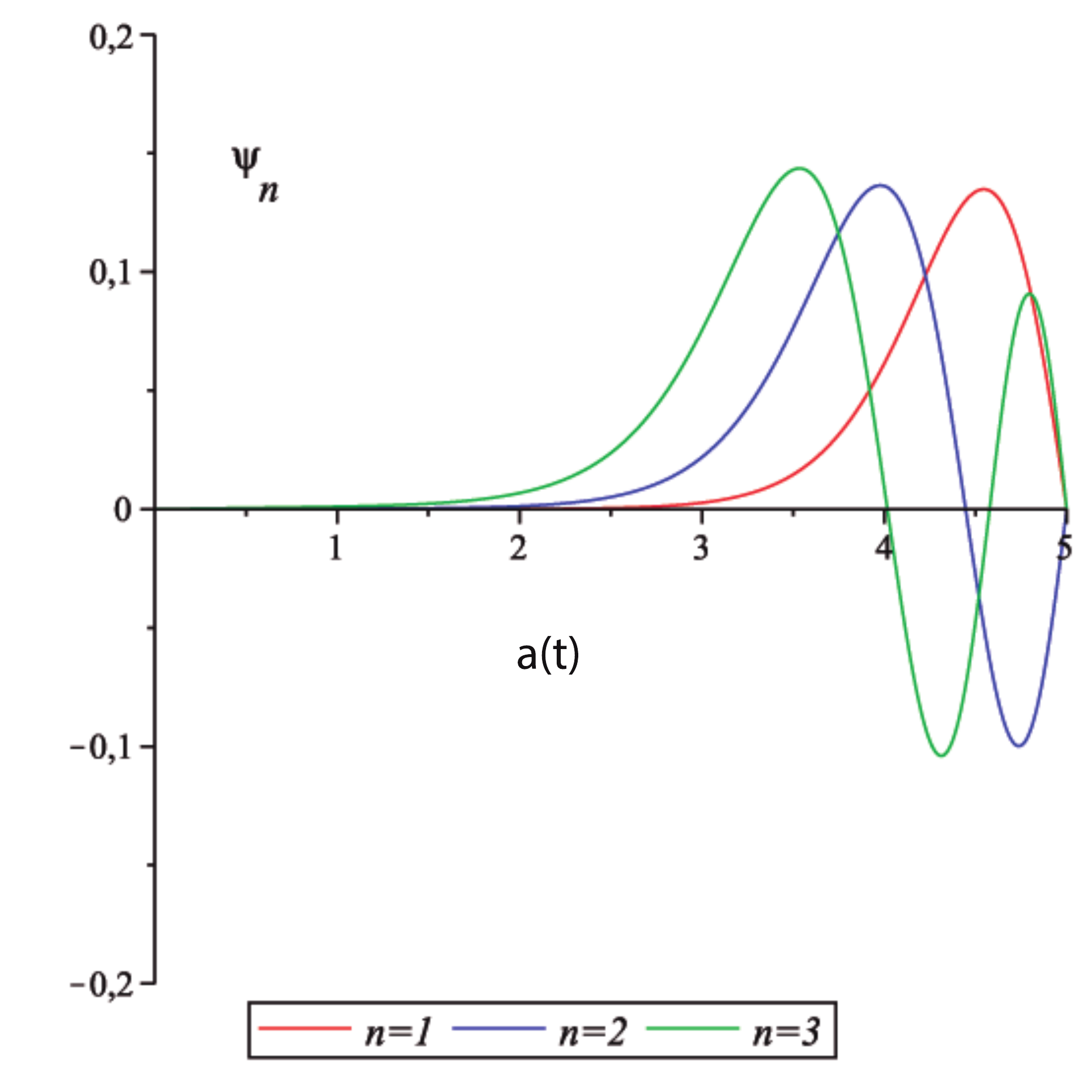}
\caption{Examples of eigenstates for the case of a domain walls fluid.}
\end{subfigure}
\begin{subfigure}{0.5\textwidth}
\includegraphics[width=0.9\linewidth, height=5cm]{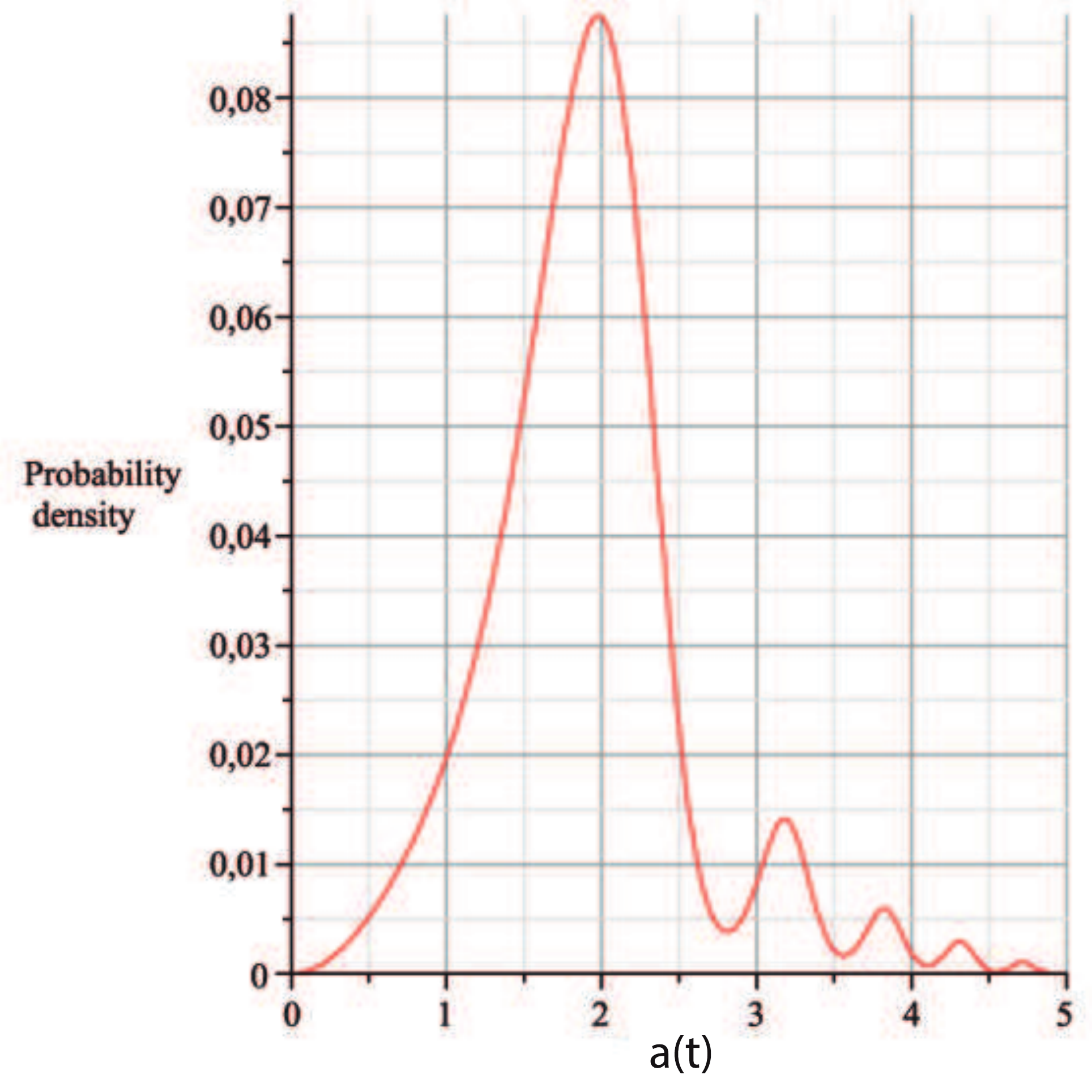} 
\caption{Initial probability density  $|\Psi(a,0)|^2$  for the domain walls fluid.}
\end{subfigure}
 \begin{subfigure}{0.5\textwidth}
\includegraphics[width=0.9\linewidth, height=5cm]{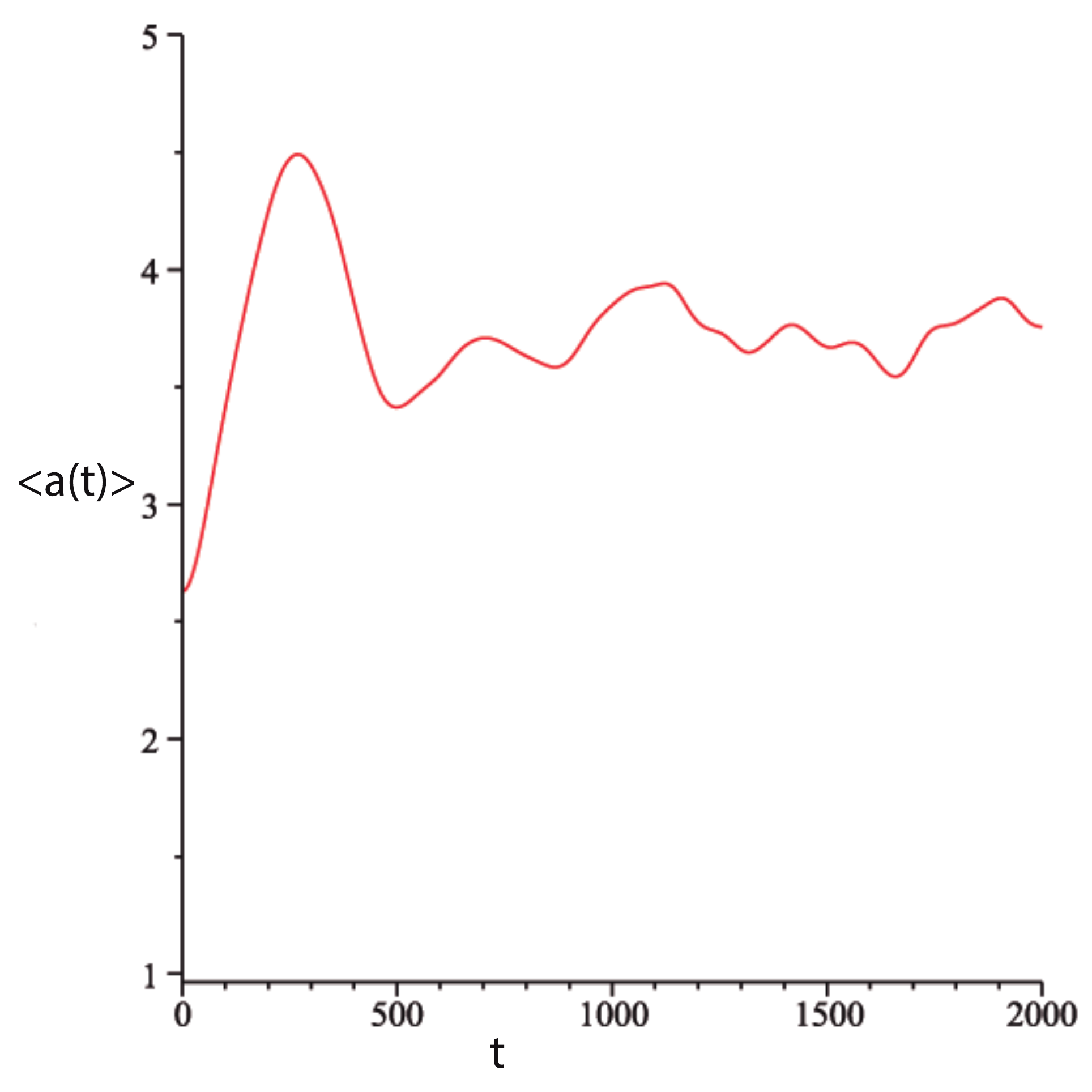} 
\caption{Expected value $\left<a\right>$, as a function of the time $t$ for the domain walls fluid.}
\end{subfigure}
\caption{Cosmological solutions for the case of a domain walls fluid.}
	\label{pd01}
\end{figure}

\begin{figure}[h!]
\includegraphics[width=0.9\linewidth, height=5cm]{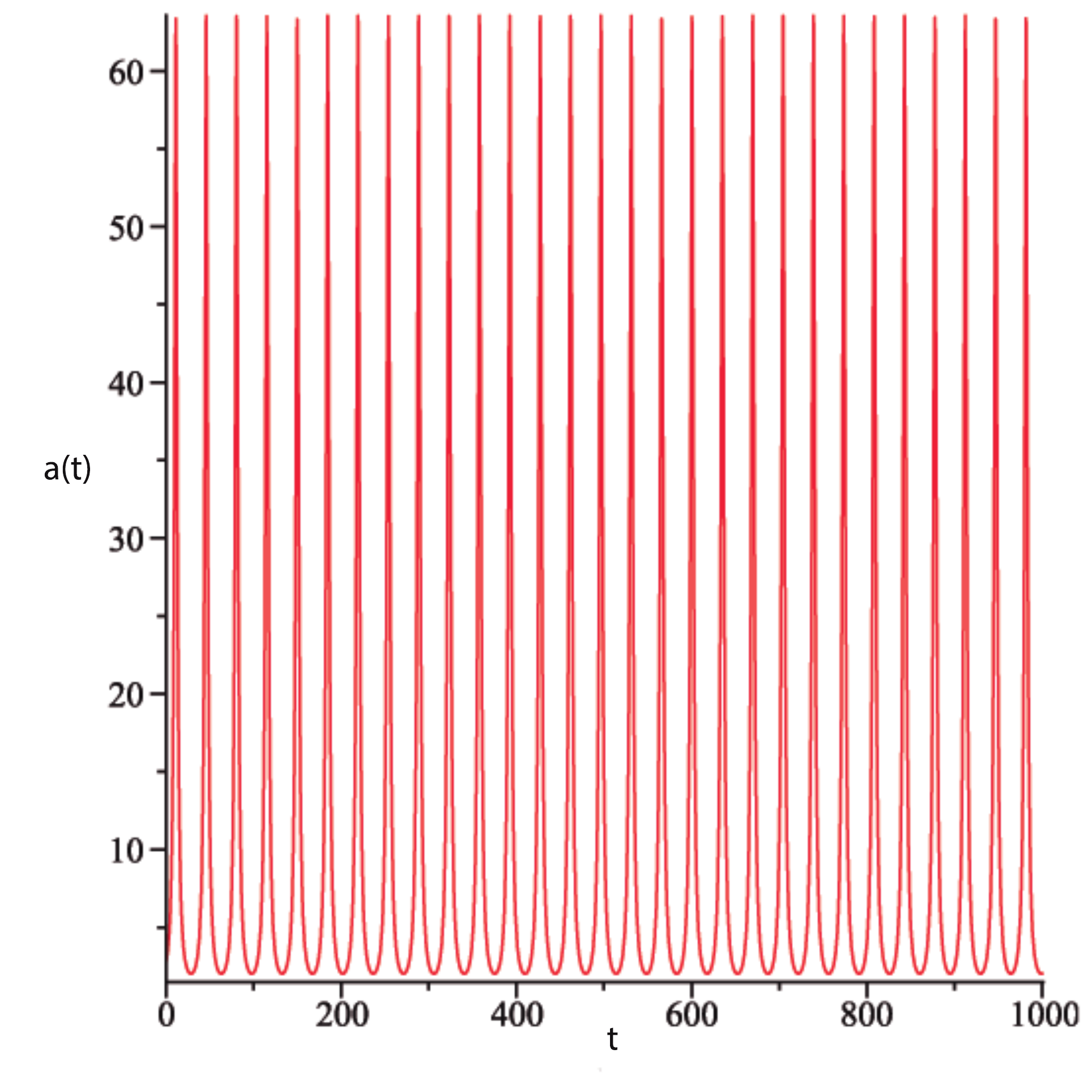} 
\caption{classical scale factor behavior for the corresponding
model to the one shown in Figure \ref{pd01}}
\label{classicaldw}
\end{figure}

Since, the 
energies $E_j$ of the others fluids, in the potential $V_{efdw}(a)$ eq. (\ref{peparedes}),
were not quantized, they have a continuum spectrum.
Again, the effective potential yields bounded states. In this case the weight function of the inner product of wave functions is $a^3$. Like in the previous case, we have applied Galerkin method and obtained approximate eigenvalues and eigenfunctions. Fig. \ref{pd01} shows the quantum behavior of this model with the time variable corresponding to the domain walls fluid.
There, we can see the expected value of the scale factor of the Universe, the first three 
eigenstates, as well as the initial probability density and the effective 
potential obtained. In this example we used: $E_{v}=-0.0017,\,\, E_{d}=1/24,\,\, E_{cs}=1/24,\,\, E_{r}=1/24,\,\, L=5$.
In Figure \ref{classicaldw}, we show the classical scale factor behavior for the corresponding
model to the one shown in Figure \ref{pd01}. In the classical model we used:
$p_{T_{v}}=-0.0017,\,\, p_{T_{d}}=1/24,\,\, p_{T_{cs}}=1/24,\,\, p_{T_{r}}=1/24$, which are identical to the corresponding energies in the quantum model.
For $p_{T_{cs}}$, we considered the mean value of the ten eigenvalues used
in order to construct the wavepacket of the quantum model: $p_{T_{dw}}=0.11195$. 
As the scale factor initial value ($a(0)$), we used the scale factor expected value at $t=0$: $a(0)=2.62901900604848$.
Finally, the initial value for the scale factor time derivative ($\dot{a}(0)$) was obtained by using the appropriate Friedmann equation
with the values given above for $p_{T_{v}}, p_{T_{d}}, p_{T_{dw}}, p_{T_{r}}, p_{T_{cs}}, a(0)$: $\dot{a}(0)=0.2504381726$.
Comparing Figures \ref{pd01}(d) and \ref{classicaldw}, we notice that the scale factor expected value is oscillating, such that, its maxima and minima values are inside the region where the classical scale factor trajectory oscillates.


\subsection{Radiation fluid $(\omega_{r}=1/3)$ }

The parameter of the equation of state of relativistic matter (radiation, i.e. photons, but also matter in the very early universe) is $\omega_{r} = 1/3$ which means that it is diluted as $\rho_{r}\propto a^{{-4}}$. In an expanding universe, the energy density decreases more quickly than the volume expansion, because radiation has momentum and, by the de Broglie hypothesis, a wavelength which is redshifted. This type of fluid is also known as hot matter, where the hot term refers to the fact that these particles have velocities equal to the speed of light $c$. They encompass not only the relativistic known elementary particles, but possibly the unknown ones (i.e. hot dark matter) \cite{cosmology,Bertone}. 
\par
In the radiation case, the Wheeler-DeWitt equation and the effective potential are given, respectively, by
\begin{equation}
\label{ewdr}
- \frac{\partial^2 \Psi(a,t)}{\partial a^2} + V_{\rm efr}(a) \Psi(a,t) = 24 i \,\frac{\partial \Psi(a,t)}{\partial t}
\end{equation}
\noindent and
\begin{equation}
V_{\rm efr}(a)= 144 a^2  -24 E_{d}\,a -24 E_{cs}\,a^2 -24 E_{dw}\,a^3 -24 E_{v}\,a^4 \,.
\label{peradiation}
\end{equation}

Since, the 
energies $E_j$ of the others fluids, in the potential $V_{efr}(a)$ eq. (\ref{peradiation}),
were not quantized, they have a continuum spectrum.
In contrast to the previous cases, Eq. (\ref{ewdr}) is a time-dependent Schr\"{o}dinger equation. The inner product 
of wave functions has weight function equal to unity.

The results are shown in Fig. \ref{fig03}. There, we can see the expected value of the scale factor of the Universe, the first three eigenstates, as well as the initial probability density and the effective 
potential obtained. In this example we used: $E_{v}=-0.0017,\,\, E_{d}=1/24,\,\, E_{cs}=1/24,\,\, E_{dw}=1/24,\,\, L=15$.
In Figure \ref{classicalr}, we show the classical scale factor behavior for the corresponding
model to the one shown in Figure \ref{fig03}. In the classical model we used:
$p_{T_{v}}=-0.0017,\,\, p_{T_{d}}=1/24,\,\, p_{T_{cs}}=1/24,\,\, p_{T_{dw}}=1/24$, which are identical to the corresponding energies in the quantum model.
For $p_{T_{r}}$, we considered the mean value of the ten eigenvalues used
in order to construct the wavepacket of the quantum model: $p_{T_{r}}=0.19$. 
As the scale factor initial value ($a(0)$), we used the scale factor expected value at $t=0$: $a(0)=9.63815017583560$.
Finally, the initial value for the scale factor time derivative ($\dot{a}(0)$) was obtained by using the appropriate Friedmann equation
with the values given above for $p_{T_{v}}, p_{T_{d}}, p_{T_{dw}}, p_{T_{r}}, p_{T_{cs}}, a(0)$: $\dot{a}(0)=0.02436255323$.
Comparing Figures \ref{fig03}(d) and \ref{classicalr}, we notice that the scale factor expected value is oscillating, such that, its maxima and minima values are inside the region where the classical scale factor trajectory oscillates.
We can verify that the temporal behavior of the expected value that is shown in Fig. \ref{fig03}(d) is similar to the behavior of the expected value in the case of a FLRW cosmological model with positive curvature, negative cosmological constant and radiation, studied in Ref. \cite{monerat5}.

\begin{figure}[h!] 
\begin{subfigure}{0.5\textwidth}
\includegraphics[width=0.9\linewidth, height=5cm]{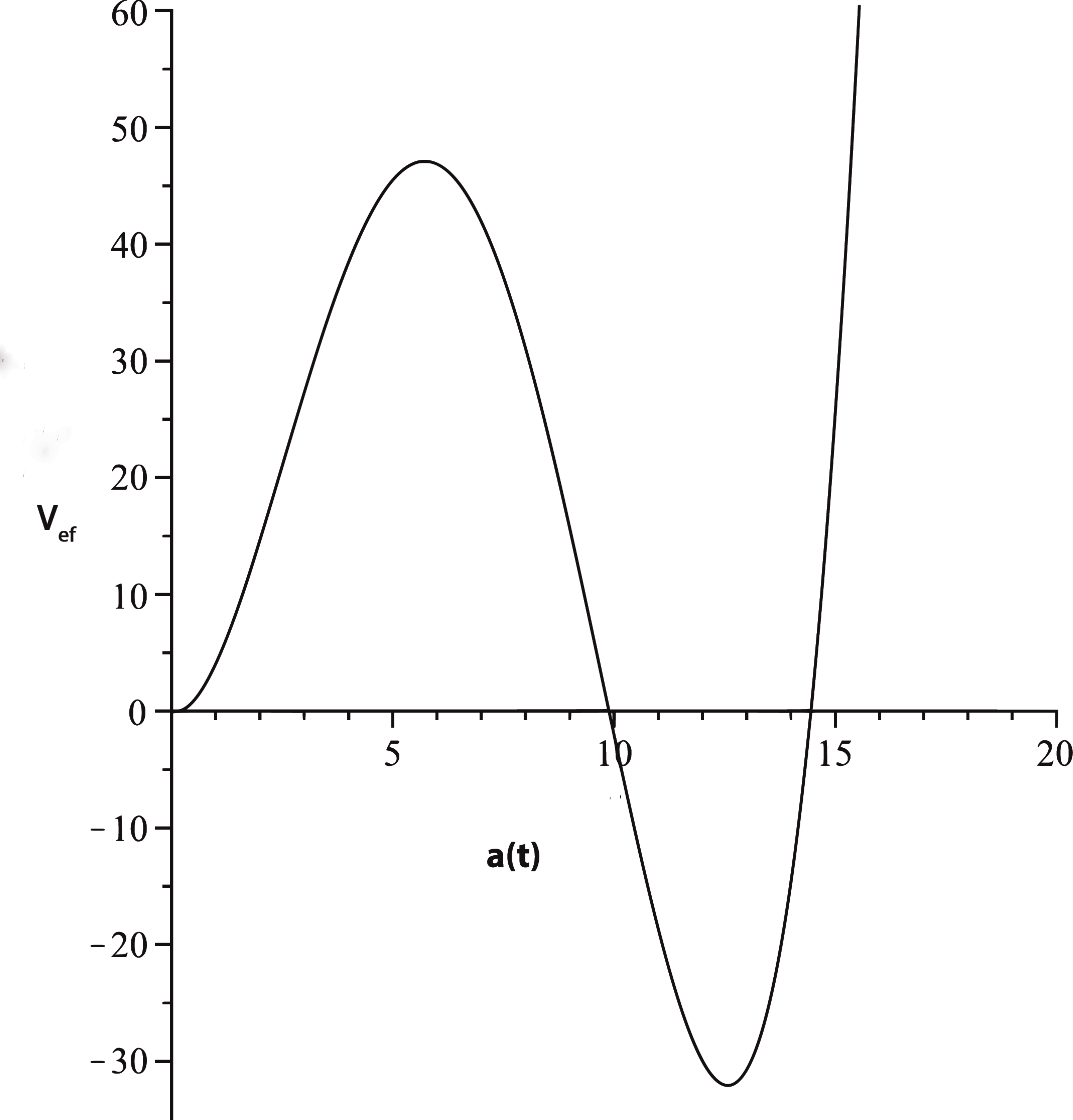} 
\caption{Effective potential for the radiation model.}
\end{subfigure}
\begin{subfigure}{0.5\textwidth}
\includegraphics[width=0.9\linewidth, height=5cm]{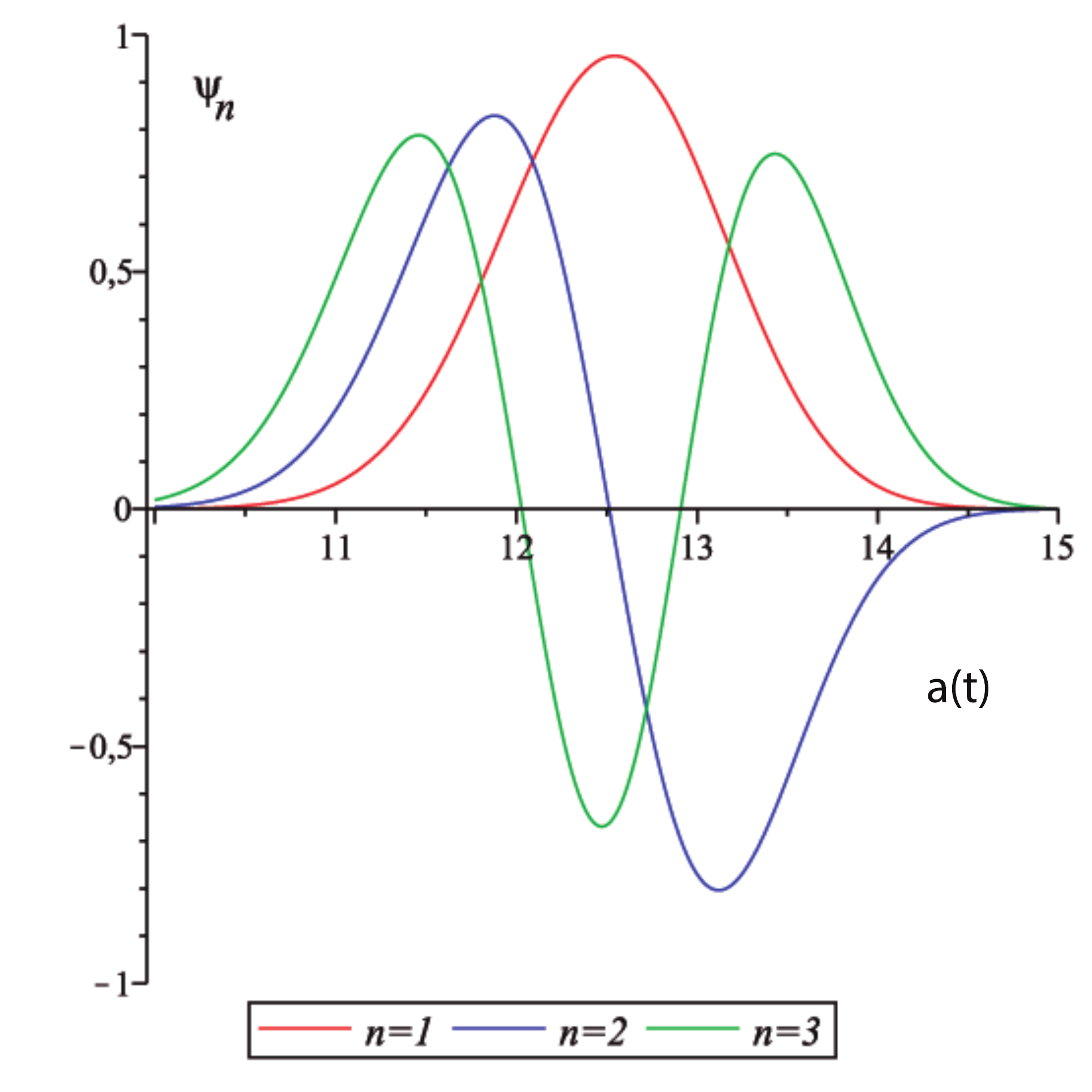}
\caption{Examples of eigenstates for the case of a radiation fluid.}
\end{subfigure}
\begin{subfigure}{0.5\textwidth}
\includegraphics[width=0.9\linewidth, height=5cm]{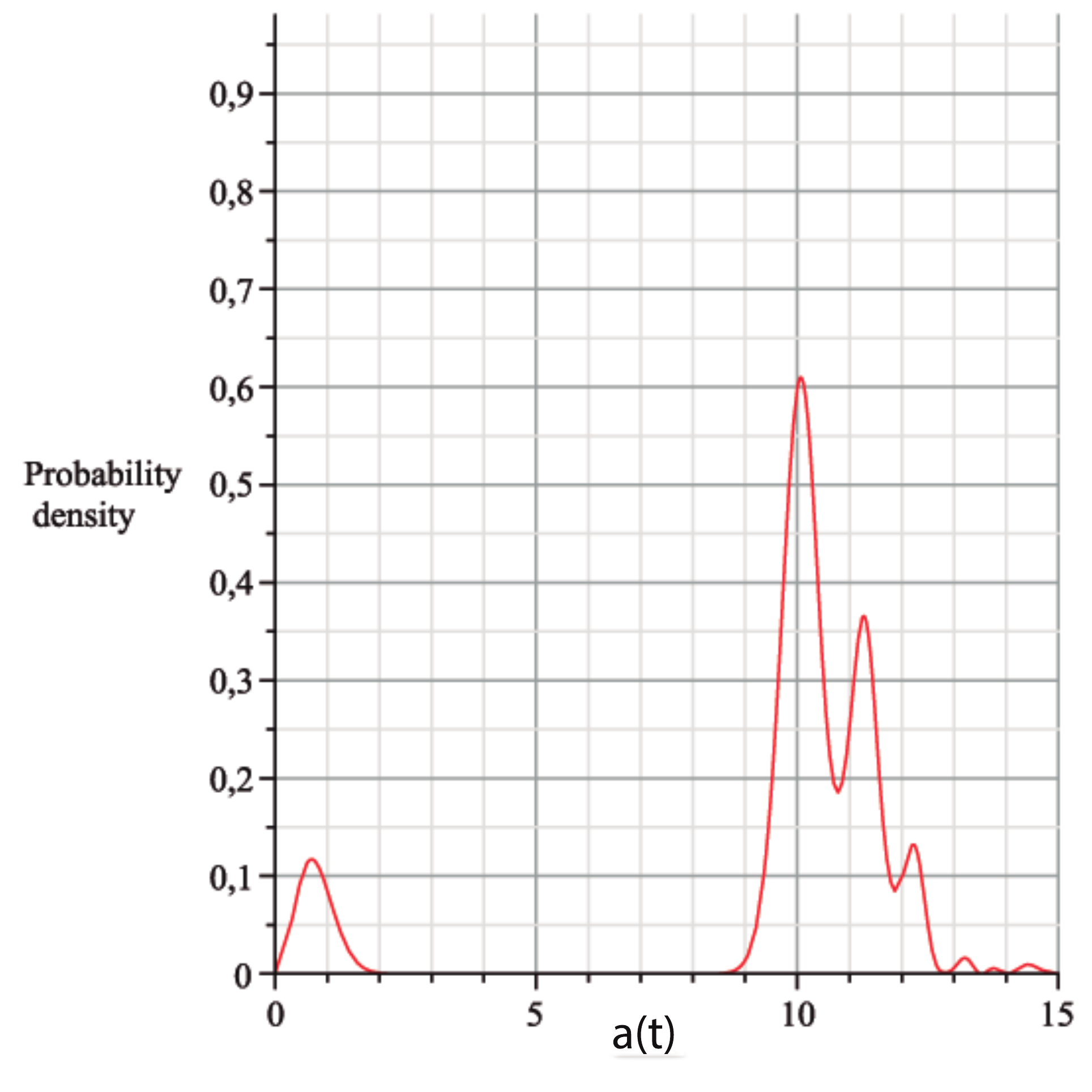} 
\caption{Initial probability density  $|\Psi(a,0)|^2$ for the radiation model.}
\end{subfigure}
\begin{subfigure}{0.5\textwidth}
\includegraphics[width=0.9\linewidth, height=5cm]{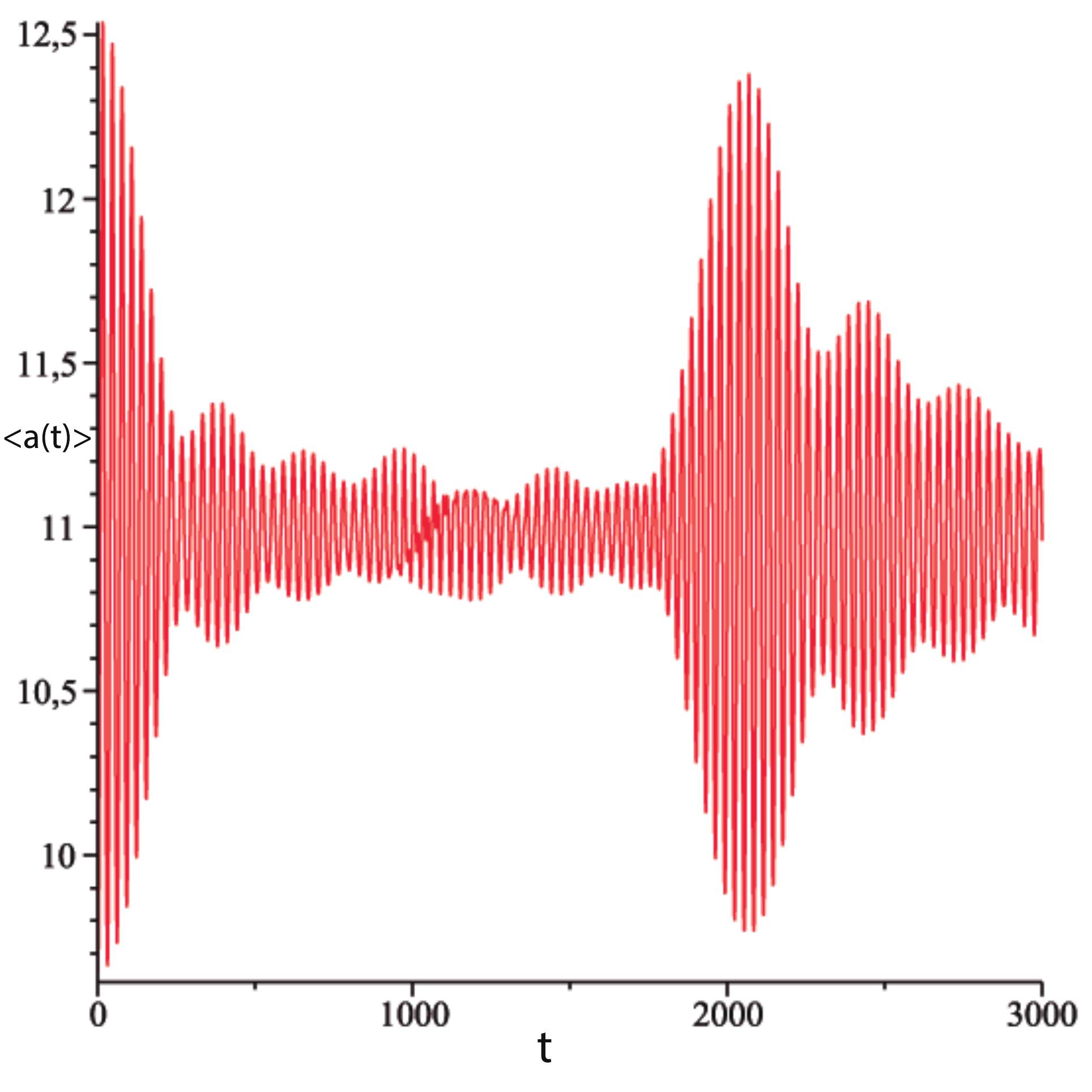} 
\caption{Expected value $\left<a\right>$, as a function of the time $t$ for the radiation fluid.}
\end{subfigure}
\label{fig03}
\caption{Cosmological solutions for the case of a radiation fluid.}
\label{fig03}
\end{figure}
\begin{figure}[h!]
\includegraphics[width=0.9\linewidth, height=5cm]{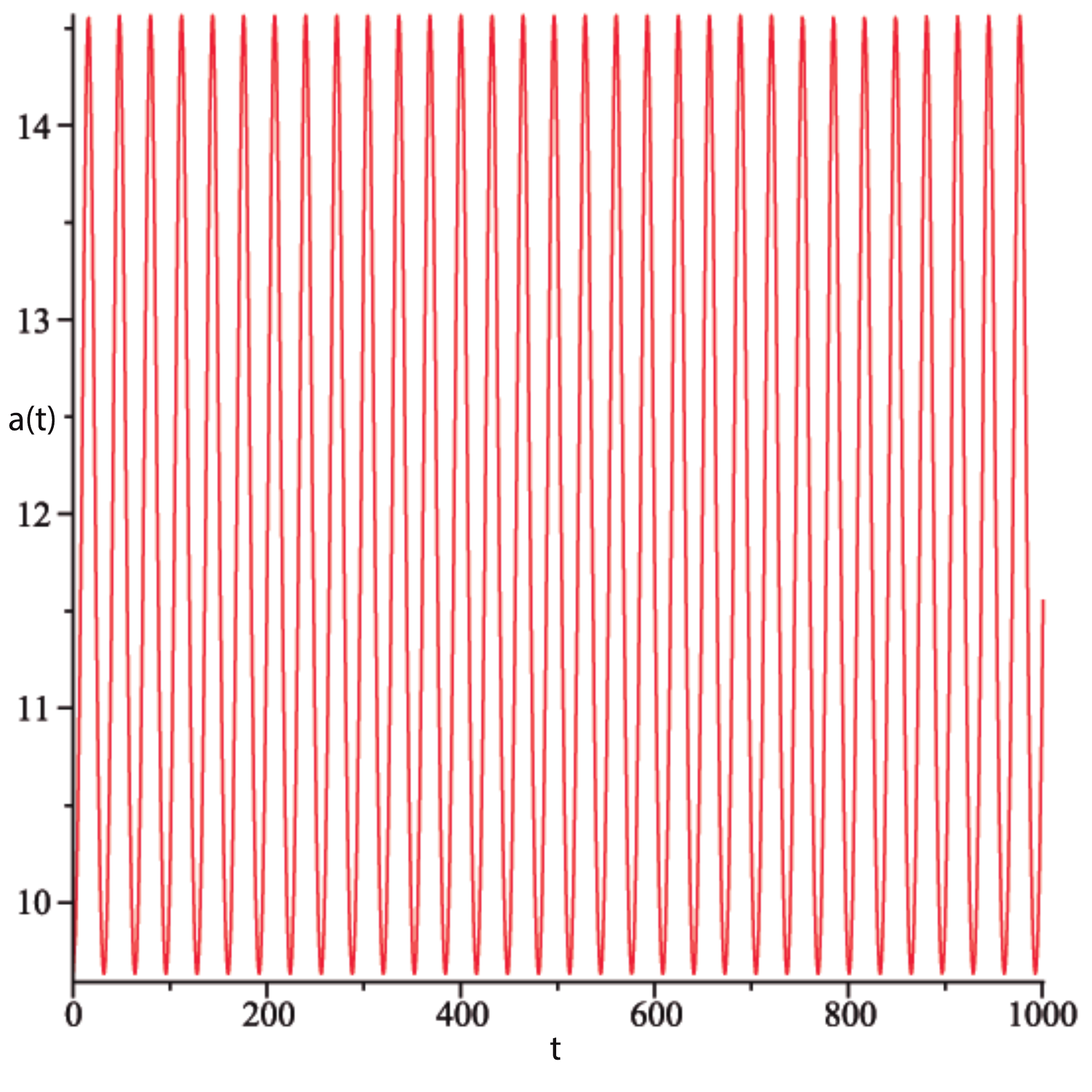} 
\caption{classical scale factor behavior for the corresponding
model to the one shown in Figure \ref{fig03}}
\label{classicalr}
\end{figure}

\subsection{Dust fluid $(\omega_{d}=0)$ }

In cosmology, dust refers to a pressureless perfect fluid, which essentially means a continuum of nonrelativistic material particles \cite{cosmology}. Galaxies behave as massive gravitationally bound entities. Since the typical separation between galaxies, $d\approx (1)$ Mpc, is much larger than the average size of a galaxy $\bar d\approx (10^{-2})$ Mpc, collision between galaxies are rare. In the hydrodynamical scenario we may therefore view the galaxies as individual particles that are massive but have no internal structure and do not collide one
another. This implies that there is no pressure between them, so it is reasonable to assume that the matter in the universe today is pressureless, with equation of state given by $p_{d} = 0$. The equation of state of ordinary non-relativistic matter is $\omega_{d} = 0$, which means that it is diluted as $\rho_{d}\propto a^{{-3}}=V^{{-1}}$, where $V$ is the volume, implying that the energy density redshifts as the volume. The presence of atoms in the universe can also be thought of as a fluid of dust. When the universe expands and consequently cools down the interactions between these elements drastically reduces and the gravitational interaction becomes dominant. Dust is also called cold matter where the adjective cold refers to the fact that particles making up this kind of matter have a kinetic energy much smaller than their mass energy \cite{Bertone}. In this case
\begin{equation}
\label{eos01}
\omega_d =\frac{p_d}{\rho_d} = \frac{mv^2_{th}}{mc^2}= \frac{v^2_{th}}{c^2}\ll 1 \quad,
\end{equation}
in which ${v_{th}}$ is the thermal velocity of particles. To a near-perfect approximation, $\omega_d = 0$, implying $\rho_d \propto a^{-3}$, in line with our simple dilution argument. Here, when we choose the dust fluid as time parameter, the Wheeler-DeWitt equation reads
\begin{equation}
\label{ewd}
- \frac{\partial^2 \Psi(a,t)}{\partial a^2} + V_{\rm efd}(a) \Psi(a,t) = 24 i \, a \,\frac{\partial \Psi(a,t)}{\partial t}\,
\end{equation}
\noindent in which
\begin{equation}
V_{\rm efd}(a)= 144 a^2  -24 E_{r} -24 E_{cs}\,a^2 -24 E_{dw}\,a^3 -24 E_{v}\,a^4 \, .
\label{pepoeira}
\end{equation}

\begin{figure}[h!] 
\begin{subfigure}{0.5\textwidth}
\includegraphics[width=0.9\linewidth, height=5cm]{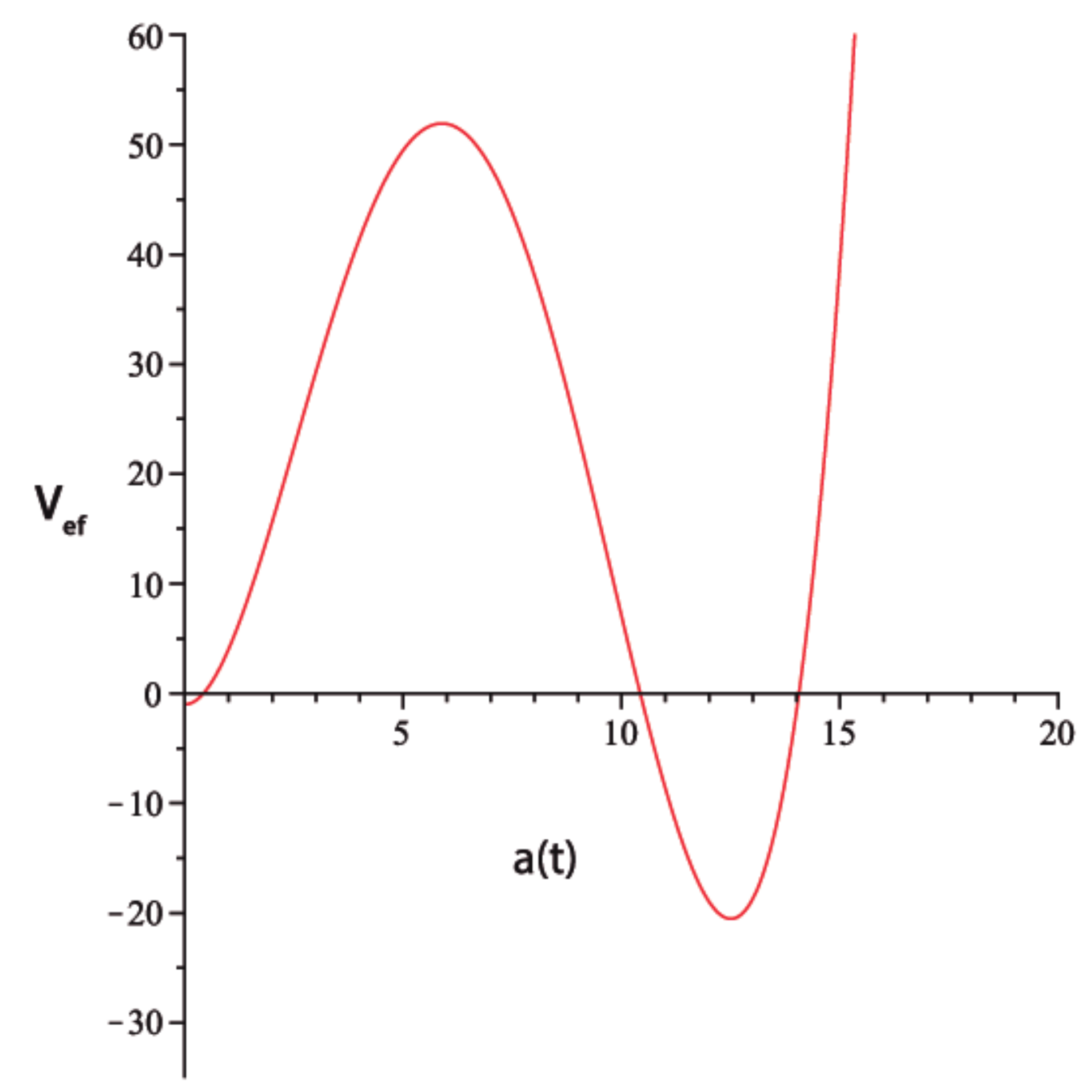} 
\caption{Effective potential for the dust model.}
\end{subfigure}
\begin{subfigure}{0.5\textwidth}
\includegraphics[width=0.9\linewidth, height=5cm]{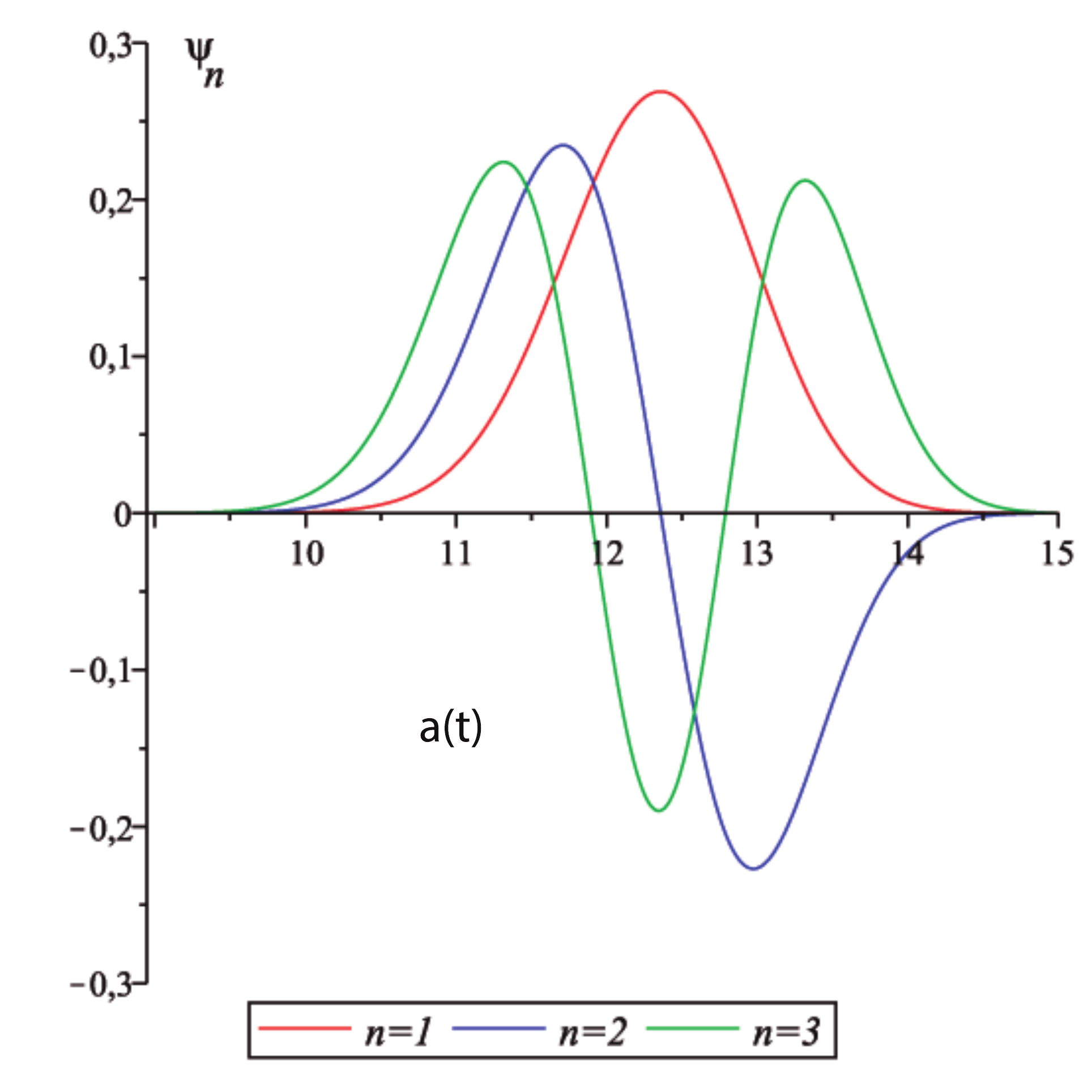}
\caption{Examples of eigenstates for the dust case.}
\end{subfigure}
\begin{subfigure}{0.5\textwidth}
\includegraphics[width=0.9\linewidth, height=5cm]{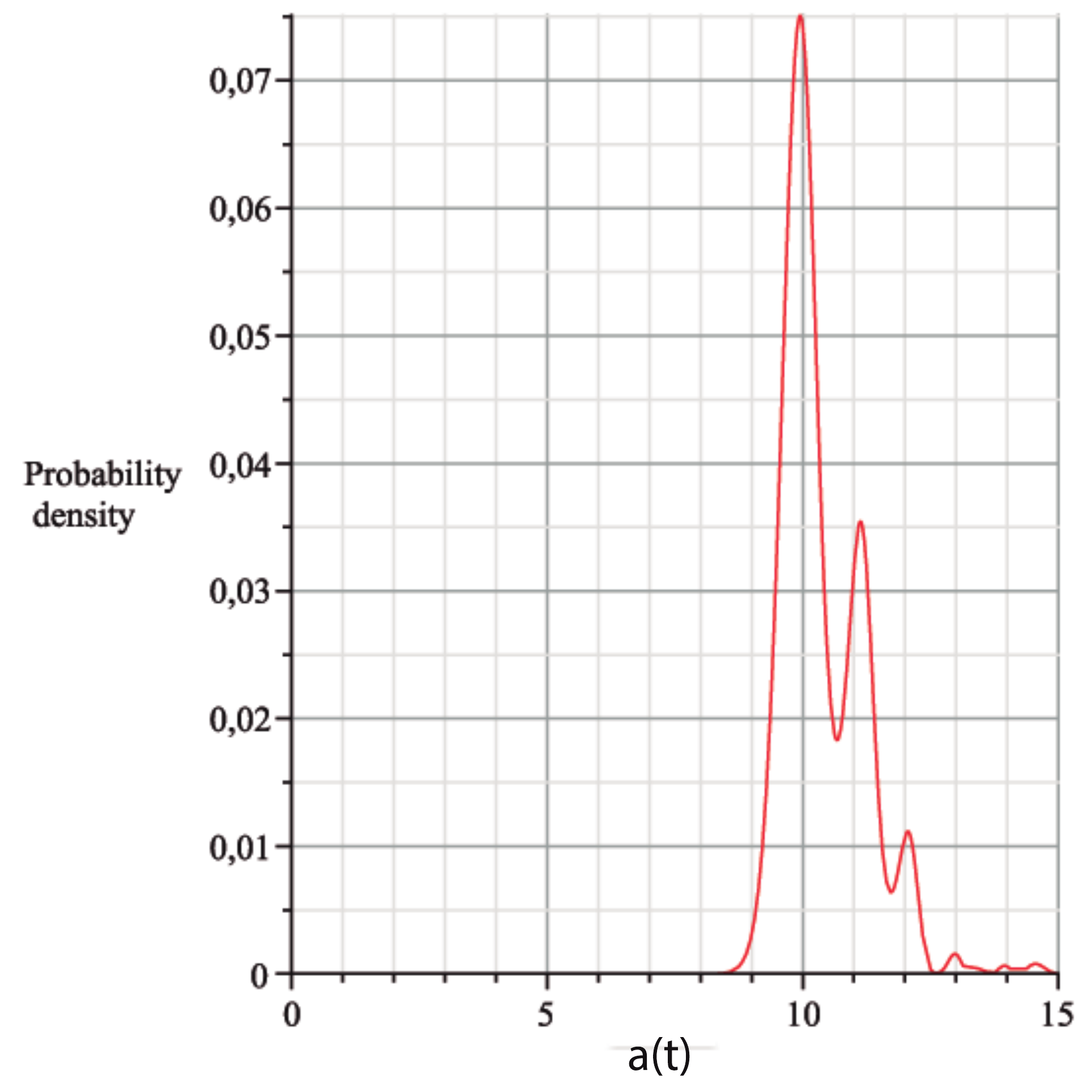} 
\caption{Initial probability density $|\Psi(a, 0)|^2$ for the dust case}
\end{subfigure}
\begin{subfigure}{0.5\textwidth}
\includegraphics[width=0.9\linewidth, height=5cm]{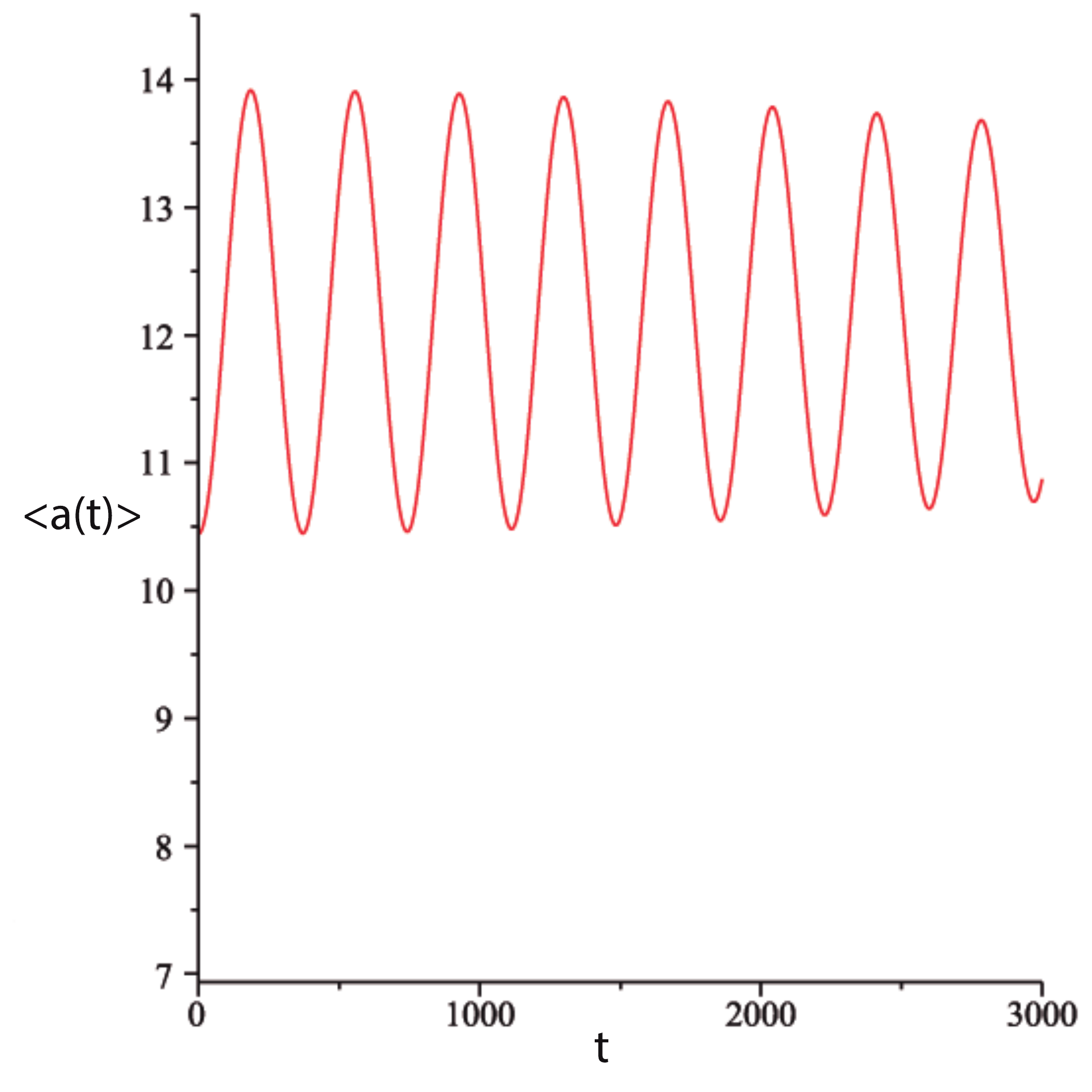} 
\caption{Expected value $\left<a\right>$, as a function of time $t$ for the dust case.}
\end{subfigure}
\caption{Cosmological solutions for the dust case.}
\label{poeirafig}
\end{figure}

\begin{figure}[h!]
\includegraphics[width=0.9\linewidth, height=5cm]{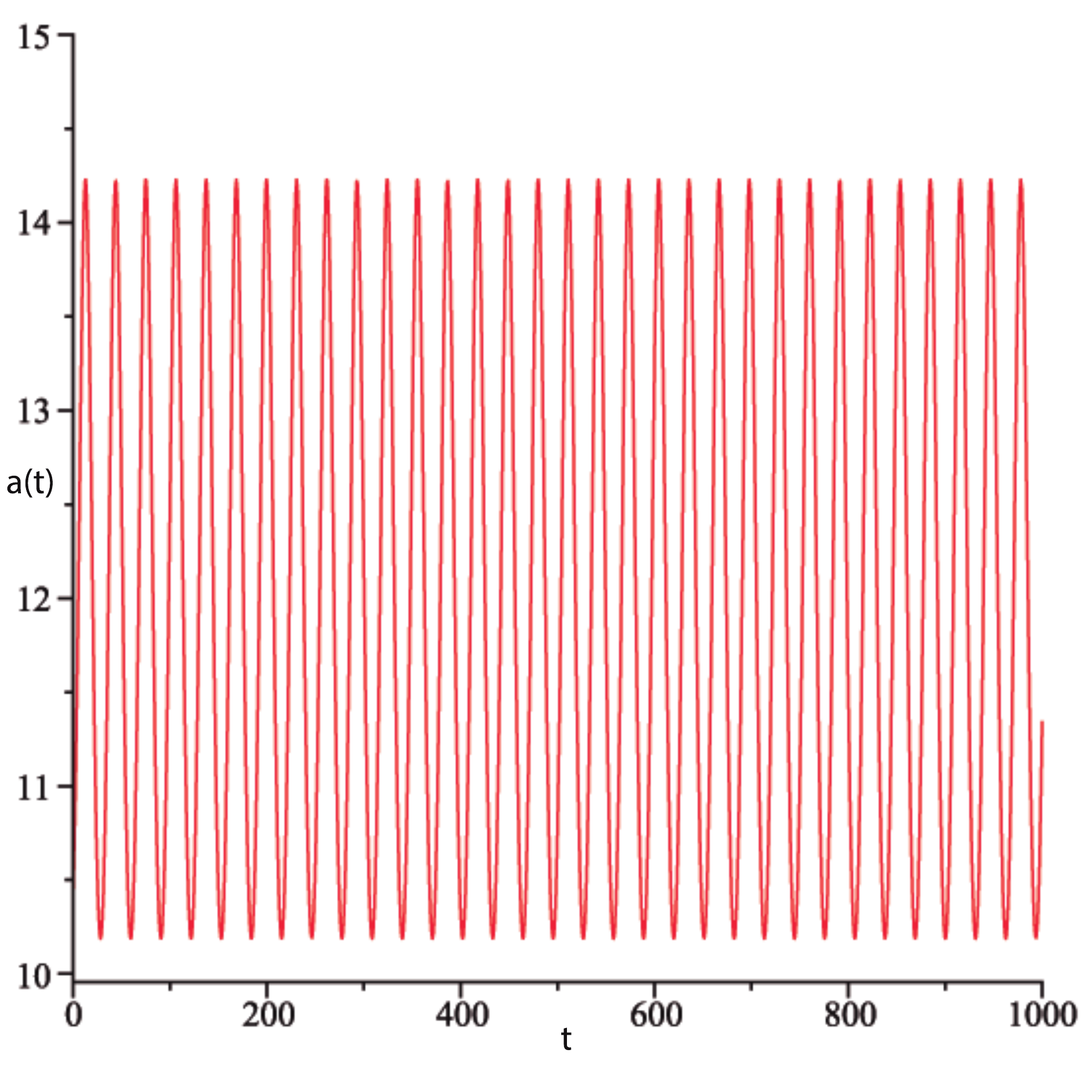} 
\caption{classical scale factor behavior for the corresponding
model to the one shown in Figure \ref{poeirafig}}
\label{classicald}
\end{figure}

Since, the 
energies $E_j$ of the others fluids, in the potential $V_{efd}(a)$ eq. (\ref{pepoeira}),
were not quantized, they have a continuum spectrum.
Once more the effective potential yields bound states, and Galerkin method can be applied for approximate solutions. In this
case, the weight function of the inner product of wave functions is $a$.
Fig. \ref{poeirafig} reveals that, as the radiation case, in the dust case the expected value of the scale factor suggests an oscillating Universe in its initial phase of evolution. 
In that figure, we can see the expected value of the scale factor of the Universe, the first three eigenstates, as well as the initial probability density and the effective 
potential obtained. In this example we used: $E_{v}=-0.0017,\,\, E_{r}=1/24,\,\, E_{cs}=1/24,\,\, E_{dw}=1/24,\,\, L=15$.
In Figure \ref{classicald}, we show the classical scale factor behavior for the corresponding
model to the one shown in Figure \ref{poeirafig}. In the classical model we used:
$p_{T_{v}}=-0.0017,\,\, p_{T_{r}}=1/24,\,\, p_{T_{cs}}=1/24,\,\, p_{T_{dw}}=1/24$, which are identical to the corresponding energies in the quantum model.
For $p_{T_{d}}$, we considered the mean value of the ten eigenvalues used
in order to construct the wavepacket of the quantum model: $p_{T_{d}}=0.015800$. 
As the scale factor initial value ($a(0)$), we used the scale factor expected value at $t=0$: $a(0)=10.4435413991694$.
Finally, the initial value for the scale factor time derivative ($\dot{a}(0)$) was obtained by using the appropriate Friedmann equation
with the values given above for $p_{T_{v}}, p_{T_{d}}, p_{T_{dw}}, p_{T_{r}}, p_{T_{cs}}, a(0)$: $\dot{a}(0)=0.1720088079$.
Comparing Figures \ref{poeirafig}(d) and \ref{classicald}, we notice that the scale factor expected value is oscillating, such that, its maxima and minima values are inside the region where the classical scale factor trajectory oscillates.


The energy spectra for cosmic strings, domain walls, radiation and dust is compared in Table \ref{tabela4} (each case corresponds to associating time with that particular fluid). For the cases of domain walls and cosmic strings, the values of $E$ grow sharply with the decrease of $\omega$. In these cases, there are no negative energies. For the choice of time in the cases of radiation and dust, the lower levels of energy E are negative as a consequence of the effective
potential, which exhibits a barrier followed by an infinite well.

In all cases examined so far $\Psi(a,t)$ is well-defined for all values of the scale factor $a$, even as $a\rightarrow 0$; moreover,
the expectation value  $\left<a\right>(t) \neq 0$ for all calculated values of $t$. Such result indicates that, at the quantum level, those models are free from singularities.

In the next section we will treat separately the fluid case of vacuum as candidate for the time variable, since there are no bound states and the Galerkin method used in the previous cases of this work can not be employed. In the vacuum scenario the potential takes the form of a barrier suggesting the possibility of the quantum tunneling process of the region classically forbidden to that allowed as provided in \cite{Atkatz}. An initial oscillating phase is not observed, which suggests the characteristic expansion of the Big Bang scenario.

\begin{table}[h!]
		\centering
{\scriptsize
\begin{tabular}{|c|c|c|c|c|}
\hline
$ $ & Domain walls           & Cosmic strings                    & Dust               & Radiation      \\ \hline
$n$ & $\omega = -\frac{2}{3}$           & $\omega = -\frac{1}{3}$   &$\omega = 0$          &$\omega = \frac{1}{3}$ \\ \hline
1   & 0.06384434922315282               & 5.741681853598941         &-0.06030416675251883  & -1.226563151753191    \\ \hline
2   & 0.07128227629787177               & 5.74305444512315          &-0.04336190951161208  & -1.009301233311391    \\ \hline
3   & 0.07902844926431654               & 5.744427027796942         &-0.02646347603383055  & -0.7950178186299972   \\ \hline
4   & 0.08744876364554739               & 5.745799612626338         &-0.009609938940970791 & -0.5838239321779358   \\ \hline
5   & 0.0967239608301227                & 5.747172309620166         &0.007199411795835328  & -0.3758150502078255  \\ \hline
6   & 0.1069800663819491                & 5.748545865051883         &0.02397290901655459   & -0.1710258210156473  \\ \hline
7   & 0.1183249043853457                & 5.749923745980198         &0.040741149343297     & 0.03063547619994838  \\ \hline
8   & 0.1308704790919839                & 5.751316611152939         &0.05757766533560688   & 0.2294195862145016   \\ \hline
9   & 0.1447406500210076                & 5.752744992427849         &0.07460265187871977   & 0.2475627099398965    \\ \hline
10  & 0.1600571130843507                & 5.754232378223624         &0.0919456155230769    & 0.4256530613952674   \\ \hline
\end{tabular}}
\caption{Eigenstates calculated by the Galerkin spectral method. Here, $n$ indexes the eigenvalues.
The value of the parameter $L$ differs in each case; here we have used $L = 15$ (for dust, radiation and cosmic strings), 
and  $L = 5$ for domain walls.}
\label{tabela4}
\end{table}

\section{Vacuum fluid $(\omega_v = -1)$}

The history of the cosmological constant and its introduction into the relativistic equations of gravitation is well known: Einstein sought to obtain a static universe but his equations provided a dynamic universe. He introduced the cosmological constant to avoid this dynamic behaviour. When Hubble showed that the universe in which we live was really dynamic and that its components were moving away from each other the use of the cosmological constant lost its meaning. But his concept returned in at least three different contexts: (i) as the vacuum energy in quantum field theories; (ii) as responsible for the cosmic inflation and  (iii) dark energy related to the
acceleration of the cosmic expansion \cite{Peebles}.

To understand the vacuum case in terms of hydrodynamics concepts and its use in cosmological scenarios we start with the Friedmann equation with a non-zero cosmological constant $\Lambda$
\begin{equation}
\label{FLRW01}
3\frac{\ddot a}{a} = {\Lambda} - 4{\pi} G(\rho + 3p)\quad.
\end{equation}

Next, we want to treat $\Lambda$ as an effective pressure and energy density as
\begin{eqnarray}
p_{\rm ef} &=& p - \frac{\Lambda}{8\pi G}\quad,\\
\rho_{\rm ef} &=& \rho + \frac{\Lambda}{8\pi G}\quad.
\end{eqnarray}

We do this because now the Friedmann equation (\ref{FLRW01}) simplifies to
\begin{equation}
3 \frac{\ddot a}{a} = -4\pi G(\rm \rho_{ef} + 3 p_{\rm ef})\quad .
\end{equation}
\noindent So now we get an effective value for $\omega$ related to our equation of state
\begin{equation}
\omega_{\rm ef} = \frac{p_{\rm ef}}{\rho_{\rm ef}} = \frac{p - \frac{\Lambda}{8\pi G}}{\rho + \frac{\Lambda}{8\pi G}}\quad,
\end{equation}
and if we assume that the cosmological constant is dominant, i.e. $p\approx 0$ and $\rho\approx 0$, we get the value of $\omega_{\rm ef}$ for vacuum fluid is $-1$.

It is easy to verify that the cosmological constant $\Lambda$ is a dimensionful parameter with units of $(length)^{-2}$. From the point of view of classical general relativity, there is no preferred choice for what the length scale defined by $\Lambda$ might be. Particle physics and quantum field theories, however, bring a different perspective to the question. Here, the existence of $\Lambda$ comes from zero point energy of fluctuation, that is zero point energy of harmonic oscillators which represent the quanta of the field. Vacuum energy has the special property that its density is a constant. By that we mean that the density of vacuum energy inside the universe does not change when you change the size of the universe. The vacuum energy density is called $\rho_0$. Now, the variable named $\Lambda$ is just $\rho_0$ multiplied by a constant.
By definition $\Lambda = (8\pi G/3)~\rho_0$. So $\Lambda$ is not a fundamentally different concept than the vacuum energy density. As we have seen above, the vacuum does correspond to an equation of state with $\omega_v = -1$. If the energy density of the vacuum is positive, the pressure is negative. If the energy density is negative, the pressure is positive. It is a characteristic of vacuum energy \cite{Peebles}.

The Wheeler-DeWitt equation for the vacuum case ($\omega_v = -1$) takes the form
\begin{equation}
-\frac{\partial^2 \Psi(a,t)}{\partial a^2} + V_{\rm efv}(a)\Psi(a,t)=24i\, a^4\, \frac{\partial \Psi(a,t)}{\partial t}\quad,
\label{WDW-vacuo}
\end{equation}
in which the effective potential is
\begin{equation}
\label{efv}
V_{\rm efv}(a)= 144 a^2 - 24 E_{d}\,a - 24 E_{r} -24 E_{cs}\,a^2 -24 E_{dw}\,a^3\quad.
\end{equation} 

\begin{figure}[!h] 
\begin{center}
\includegraphics[height=5.5cm,width=7cm]{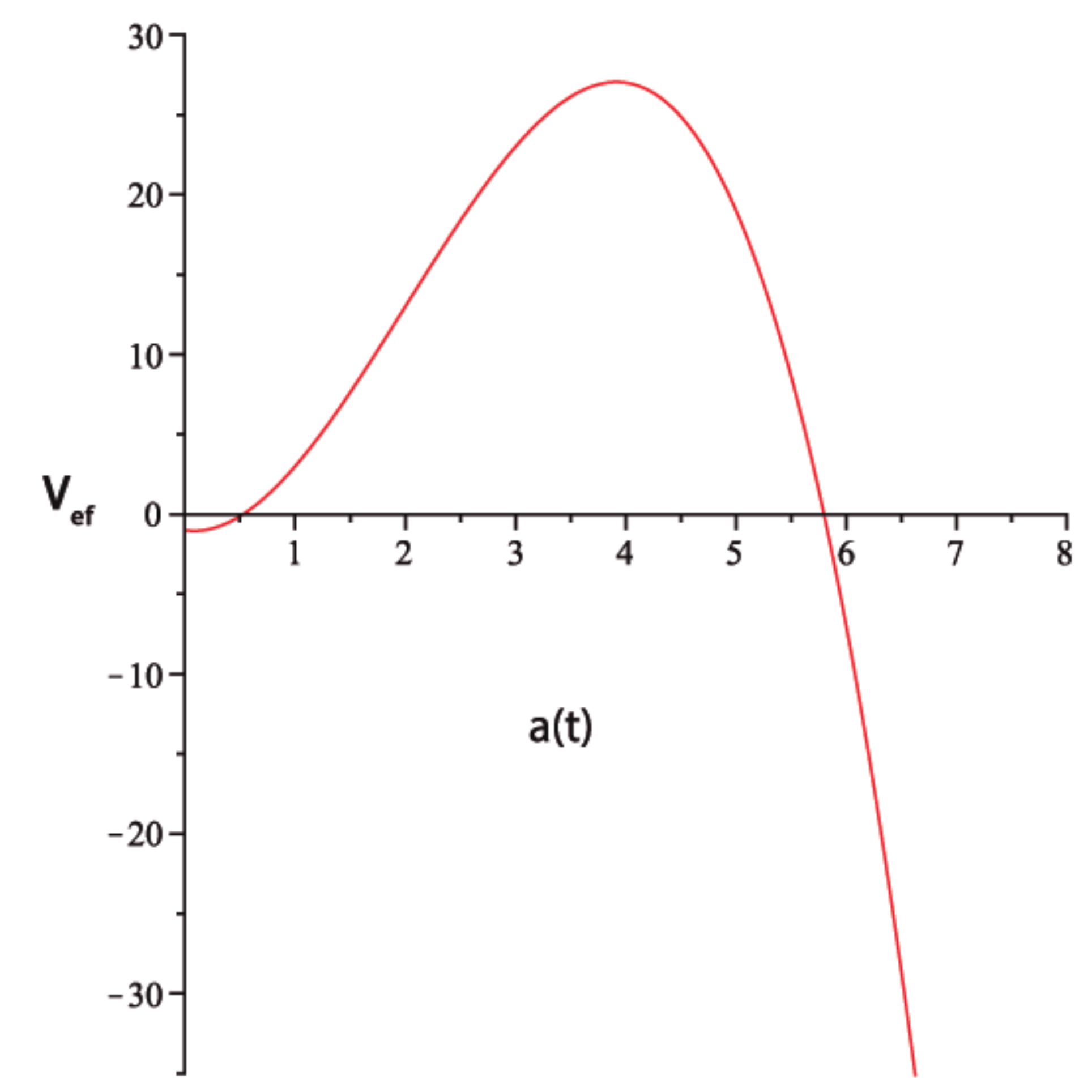}%
\caption{Effective potential $V_{\rm ef}(a)$ for the vacuum case ($\omega_v=-1$). Here we consider $p_{T_{v}}=-0.0017,\,\, p_{T_{d}}=1/24,\,\, p_{T_{dw}}=1/24,\,\, p_{T_{cs}}=5.75,\,\, p_{T_{r}}=1/24$.}
	\label{potencial-vacuo01}
\end{center}
\end{figure}

Since, the 
energies $E_j$ of the others fluids, in the potential $V_{efv}(a)$ eq. (\ref{efv}),
were not quantized, they have a continuum spectrum.
The actual potential now describes a small well followed by a potential barrier, as shown in the Figure \ref{potencial-vacuo01}. It is observed that the vacuum case does not have a bound state structure. Unlike the other fluids candidates for the role of time, the model presents a continuous energy spectrum, and can not be analyzed by the Galerkin spectral method, as done in previous cases. The use of Galerkin's method in this case could be understood as an approximation of a more general situation. Thus, the finite difference method is employed in the Crank-Nicolson scheme \cite{Cranck} because of its known stability.

Eq. (\ref{WDW-vacuo}) is not in the form of a time-dependent Schrödinger equation, since the time derivative is multiplied by a factor $a^4$. Because of this, we apply a canonical transformation to the case $\omega_v = -1$  \cite{monerat15}
\begin{equation}
a = \sqrt[3]{3x}\,\,\, ;\,\,\, p_a = p_x a^2,
\label{transformacao-canonica}
\end{equation}
so that the Hamiltonian takes the form
\begin{equation}
H=-\frac{1}{24}\, {p_{x}}^2-\frac{2\sqrt[3]{3}}{\sqrt[3]{x^{2}}}+\frac{1}{9}\frac{\sqrt[3]{9}{p_{T}}_{r}}{\sqrt[3]{x^{4}}}+\frac{{p_{T}}_{d}}{3x}+\frac{\sqrt[3]{3}{p_{T}}_{cs}}{3\sqrt[3]{x^{2}}}+\frac{\sqrt[3]{9}{p_{T}}_{dw}}{3\sqrt[3]{x}}+{p_{T}}_{v}\,\,,
\label{hnew}
\end{equation}
thus recasting Eq. (\ref{WDW-vacuo}) as a legitimate time-dependent Schrödinger equation
\begin{equation}
-\frac{\partial^2 \Psi(x,t)}{\partial x^2}+V_{\rm efv}(x)\Psi(x,t)=24i\, \frac{\partial \Psi(x,t)}{\partial t}\quad,
\label{schrodinger-vacuo}
\end{equation}
with an effective potential in the form
\begin{equation}
V_{\rm efv}(x)=\frac{2\sqrt[3]{3}}{\sqrt[3]{x^2}}-\frac{\sqrt[3]{9}E_r}{{9\sqrt[3]{x^4}}}-\frac{E_d}{3x}
-\frac{\sqrt[3]{3}E_{cs}}{3\sqrt[3]{x^2}}-\frac{\sqrt[3]{9}E_{dw}}{3\sqrt[3]{x}}\quad.
\label{Vx-vacuo-x}
\end{equation}

\begin{figure}[!h] 
\begin{center}
\includegraphics[height=5.0cm,width=6cm]{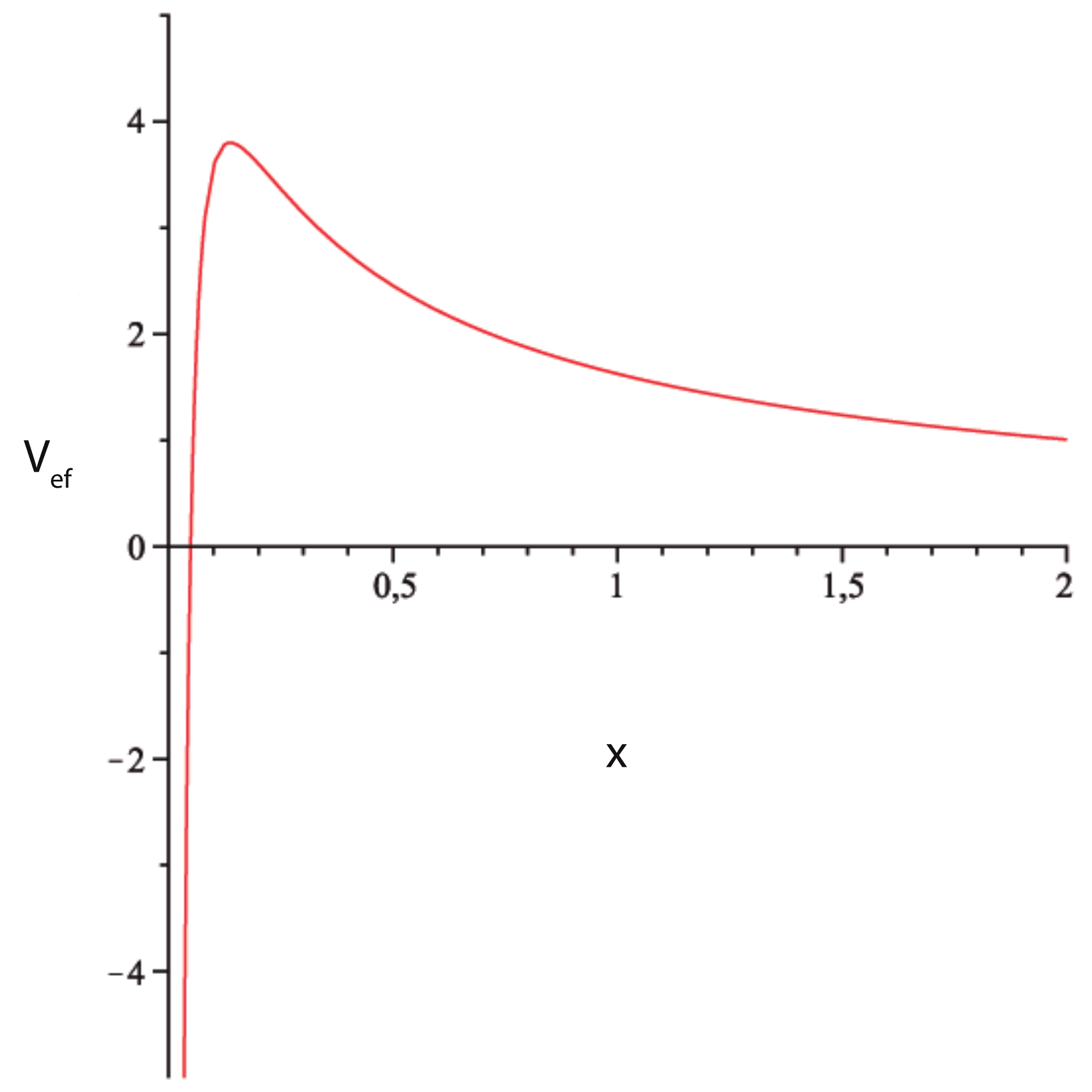}%
\caption{Effective potential $V_{\rm ef}(x)$ for the vacuum case ($\omega_{v}=-1$). Here we consider $k=1,\,\, p_{T_{v}}=-0.0017,\,\, p_{T_{d}}=1/24,\,\, p_{T_{dw}}=1/24,\,\, p_{T_{cs}}=5.75,\,\, p_{T_{r}}=1/24$.}
\label{potencial-vacuo-x}
\end{center}
\end{figure}

The effective potential in the $x$ variable will take the form of a small potential barrier as shown in Fig. \ref{potencial-vacuo-x}.

As initial condition have chosen the normalized wave function
\begin{equation}
\Psi(x,0)=\left(\frac{8192E_{m}^3}{\pi}\right)^{1/4}x\, e^{-4{E_{m}}\,\,{x^2}}\quad ,
\label{condicao-inicial}
\end{equation}
which depends on the average kinetic energy of the initial packet ($E_{\rm m}$) and satisfies the required boundary conditions $\Psi(0, t) = \Psi(\infty, t) = 0$. 

\begin{figure}[!h] 
\begin{center}
\includegraphics[height=5.0cm,width=6cm]{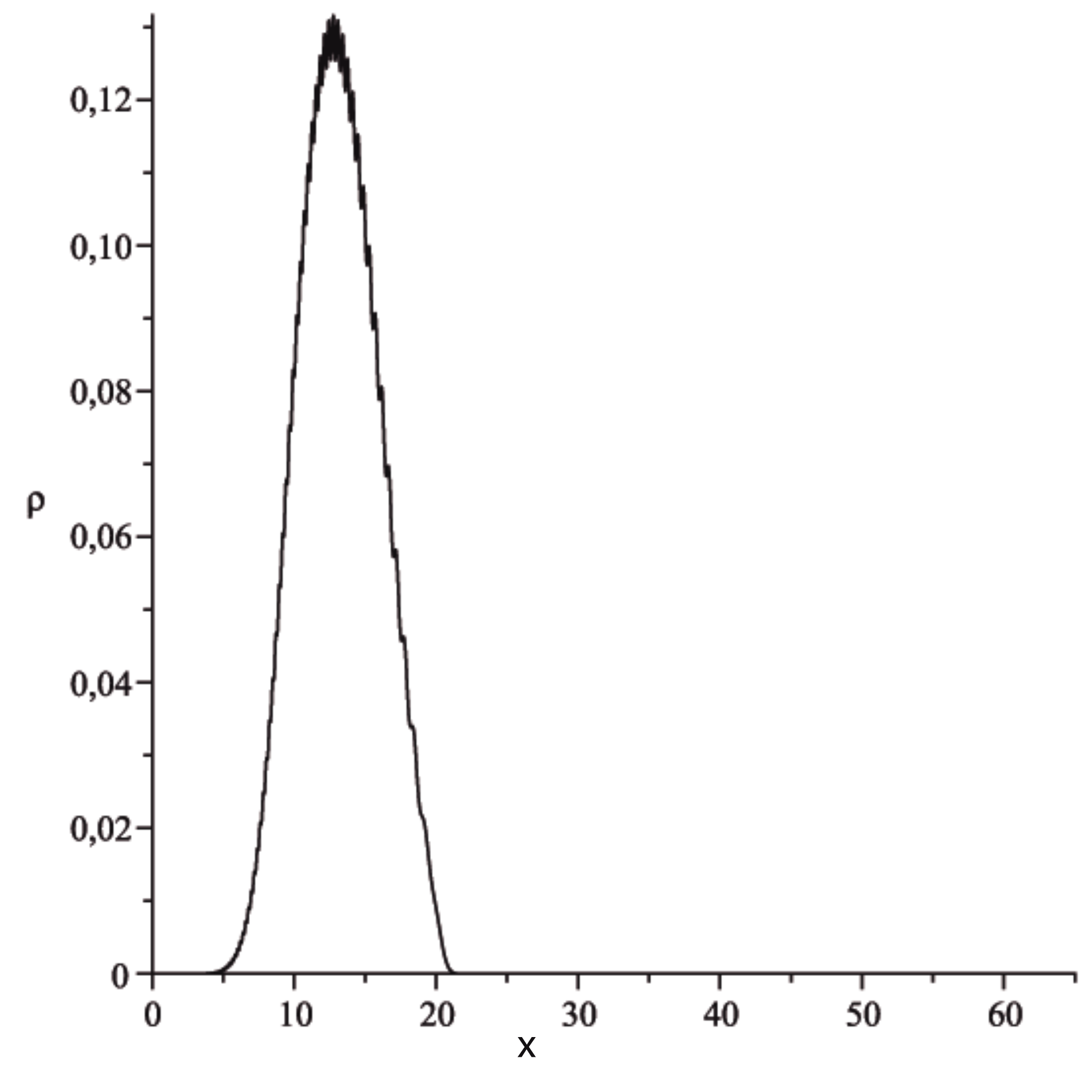}%
\caption{$|\Psi(x,t_{\rm max})|^2=\rho$ for $E_{\rm m}=3.5$ and $t_{\rm f}=10$  when $\Psi$ reaches the numerical infinity in $x_{\rm f}=65$.}
\label{Pacote-de-onda-x}
\end{center}
\end{figure}

Through a computational routine we implemented the Crank-Nicolson method for solving the Eq. (\ref{schrodinger-vacuo}). In all the analyzed cases it was possible to obtain a well-defined wave packet in the whole space, even with the degenerate 3-sphere. The evolution of the wave function was carried out in a reticulate of 4500 points. We considered the
numerical spatial infinity at $x_ {\rm f} = 65$ and let the wave
function evolve from $t=0$ to $t_{\rm f} = 10$, with time step $dt =
0.05$, from the initial condition (\ref{condicao-inicial}) with
$E_{\rm m }= 3.5$. Fig. \ref{Pacote-de-onda-x} shows the resulting wave packet at the instant of
time it reaches the numerical spatial infinity. The wave has
tunneled completely, indicating that the universe may arise classically
to the right of the potential barrier.

The case of the choice of vacuum fluid for the role of time differs from other cases in certain respects. Here the energy spectrum is not discrete, providing the possibility for the universe to emerge from its quantum to classical phase as long as the wave function of the universe goes through a small potential barrier. If this happens, the universe will classically appear to the right of the potential barrier and from this moment on its dynamics will be governed by Hamilton's equations.

The numerical method employed in this case is the finite difference method in the so-called Crank-Nicolson scheme. This method has been applied in the quantization of cosmological models with other material contents \cite{monerat11,Barros, monerat12}, proving effective in the search of solutions for the Wheeler-DeWitt equation. The mechanism for this is the so-called quantum tunneling, proposed by Vilenkin \cite{Vilenkin}, in which the universe can arise, as a classical system, from nothing.

\subsection{The quantum tunneling process}

Here we will investigate whether the quantum tunneling mechanism may be responsible for the birth of the universe. According to the effective potential described by Eq. (\ref{pecordas}), it has a local maximum point at $x_{\rm max} = 0.1373356097$, producing at that point a barrier with a maximum height given by $V_{\rm ef} (x_{\rm max}) = 3.803450769$. The tunneling probabilities (TP) will be calculated as proposed by \cite{Barros}
\begin{equation}
TP=\frac{\int_{x_{\rm 2}}^{x_{\rm f}}|\Psi(x,t_{\rm f})|^2dx}{\int_{0}^{x_{\rm f}}|\Psi(x,t_{\rm f})|^2dx}\quad,
\label{probabilidade-x}
\end{equation}
where $x_{\rm 2}$ corresponds to the return point to the right of the potential barrier. We also compared the results with those obtained by the WKB approximation \cite{Merzbacher}, using the expression
\begin{equation}
TP_{\rm WKB}=\frac{4}{\left(2\theta+\frac{1}{2\theta}\right)^2}\quad,
\label{probabilidade-WKB}
\end{equation}
in which
\begin{equation}
\theta=e^{\int_{x_{\rm 1}}^{x_{\rm 2}}\,dx\,\sqrt{V_{\rm ef}(x)-E}}\quad.
\label{teta-WKB}
\end{equation}
Here, $x_{\rm 1}$ is the return point to the left of the potential barrier, and the potential is given by Eq. (\ref{Vx-vacuo-x}).

Considering the energies of the system studied here as $E_{\rm m} < V_{\rm ef}(x_{\rm max})$, we have calculated the tunneling probabilities for different energy values shown in Table \ref{tabela5}.

\begin{table}[h!]
	\centering
{\scriptsize
\begin{tabular}{|c|c|c|c|c|}
\hline
$E_m$ & $TP$                    & $x_{1}$                & $x_{2}$       &$TP_{WKB}$      \\ \hline
3.5   & 1.00000                 & 0.09528990533          & 0.2223132646  & 0.5991935224   \\ \hline
3.4   & 1.00000                 & 0.09073619640          & 0.2430827355  & 0.5842817096   \\ \hline
3.3   & 0.999998                & 0.08701059396          & 0.2642240282  & 0.5685829292   \\ \hline
3.2   & 0.999999                & 0.08385282248          & 0.2861226370  & 0.5520613188   \\ \hline
3.1   & 1.00000                 & 0.08111133256          & 0.3090725728  & 0.5457093732   \\ \hline
3.0   & 1.000000                & 0.07868932583          & 0.3333333333  & 0.5164139412   \\ \hline
2.9   & 1.000000                & 0.07652086195          & 0.3591574258  & 0.5013754900   \\ \hline
2.8   & 1.000000                & 0.07455883112          & 0.3868068223  & 0.4770995536   \\ \hline
2.7   & 0.999999                & 0.07276831784          & 0.4165650298  & 0.4560138864   \\ \hline
2.6   & 1.000000                & 0.07112267672          & 0.4487477290  & 0.4339635672   \\ \hline
2.5   & 1.000000                & 0.06960108277          & 0.4837135831  & 0.4109539392  \\ \hline
2.4   & 0.999999                & 0.06818693370          & 0.5218763201  & 0.3870059309 \\ \hline
2.3   & 0.999999                & 0.06686676959          & 0.5637190778  & 0.3621599683 \\ \hline
2.2   & 0.999999                & 0.06562951995          & 0.6098120891  & 0.3364804538 \\ \hline
2.1   & 0.999999                & 0.06446596519          & 0.6608350467  & 0.3100607607 \\ \hline 
2.0   & 0.999999                & 0.06336834280          & 0.7176059329  & 0.2830286259 \\ \hline
1.9   & 0.999999                & 0.06233005366          & 0.7811187887  & 0.2555516854 \\ \hline
1.8   & 1.00000                 & 0.06134543928          & 0.8525939550  & 0.2278427446  \\ \hline 
1.7   & 0.999999                & 0.06040961020          & 0.9335459317  & 0.2001641378  \\ \hline
1.6   & 0.999999                & 0.05951831208          & 1.025876525   & 0.1728302565  \\ \hline
1.5   & 1.000000                & 0.05866781987          & 1.132004949   & 0.1462069682   \\ \hline
1.4   & 0.999999                & 0.05785485318          & 1.255053066   & 0.1207063046    \\ \hline
1.3   & 1.000000                & 0.05707650812          & 1.399114823   & 0.09677447176     \\ \hline
1.2   & 0.999999                & 0.05633020162          & 1.569657736   & 0.07487111476    \\ \hline
1.1   & 0.999999                & 0.05561362584          & 1.774137769   & 0.05543803340    \\ \hline
1.0   & 0.999999                & 0.05492471039          & 2.022971269   & 0.03885653625    \\ \hline
0.9   & 0.999997                & 0.05426159080          & 2.331128975   & 0.02539474830     \\ \hline
0.8   & 0.999984                & 0.05362258211          & 2.720866454   & 0.01514995141    \\ \hline
0.7   & 0.999863                & 0.05300615659          & 3.226652346   & 0.007996814912 \\ \hline
0.6   & 0.998468                & 0.05241092476          & 3.904652676   & 0.003559922089 \\ \hline
0.5   & 0.980635                & 0.05183561924          & 4.852518831   & 0.001236139969 \\ \hline
0.4   & 0.7988510               & 0.05127908093          & 6.255258648   & 0.0002918846693 \\ \hline
0.3   & 0.1814090               & 0.05074024699          & 8.508068111   & 0.00003586569855 \\ \hline
0.2   & 0.00125519              & 0.05021814054          & 12.62326258   & 0.000001246406622 \\ \hline
0.1   & $4.33193\cdot 10^{-12}$ & 0.04971186165          & 22.16122344   & $1.988733444\cdot 10^{-9}$ \\ \hline
\end{tabular}}
\caption{Tunneling probabilities (TP) for parameters $N=4500$ (spatial discretization), $dt=0.05$, $x_{\rm f}=65$. Here $t_{\rm f}=10$. In $x_{\rm max}=0.1373356097$ we have  $V_{\rm ef}(x_{\rm max}) = 3.803450769$. }
\label{tabela5}
\end{table}

When we compare the  tunneling probabilities $TP$ calculated for this model, we observe that these are in general much larger than those obtained in the FLRW quantum models with positive curvature and (i) with  cosmological constant and radiation \cite{monerat12}, and (ii) with the Chaplygin gas and radiation \cite{monerat11} models. This should happen due to the fact that in this work the height of the potential barrier is much smaller. Denoting the maximum potential energy of $E_{\rm max}$, the Table \ref{tabela5} shows us that the tunneling probability is $TP\approx 1$, indicating that the wave function crosses the barrier completely for energies in the range $1 \leq E_{\rm m} \leq E_{\rm max}$.

\subsection{The expected value for the scale factor of the universe}

Now, we can analyze how this quantum model predicts the behavior of the scale factor of the universe in this epoch. Using the many-worlds interpretation, we can calculate the expected value of the scale factor defined by
\begin{equation}
\left<x\right>(t)=\frac{\int_{0}^{x_{\rm f}}\,x\,|\Psi(x,t)|^2 dx}{\int_{0}^{x_{\rm f}}|\Psi(x,t)|^2 dx}\quad.
\label{xmedio}
\end{equation}

\begin{figure}[!h] 
\begin{center}
\includegraphics[height=5.5cm,width=7cm]{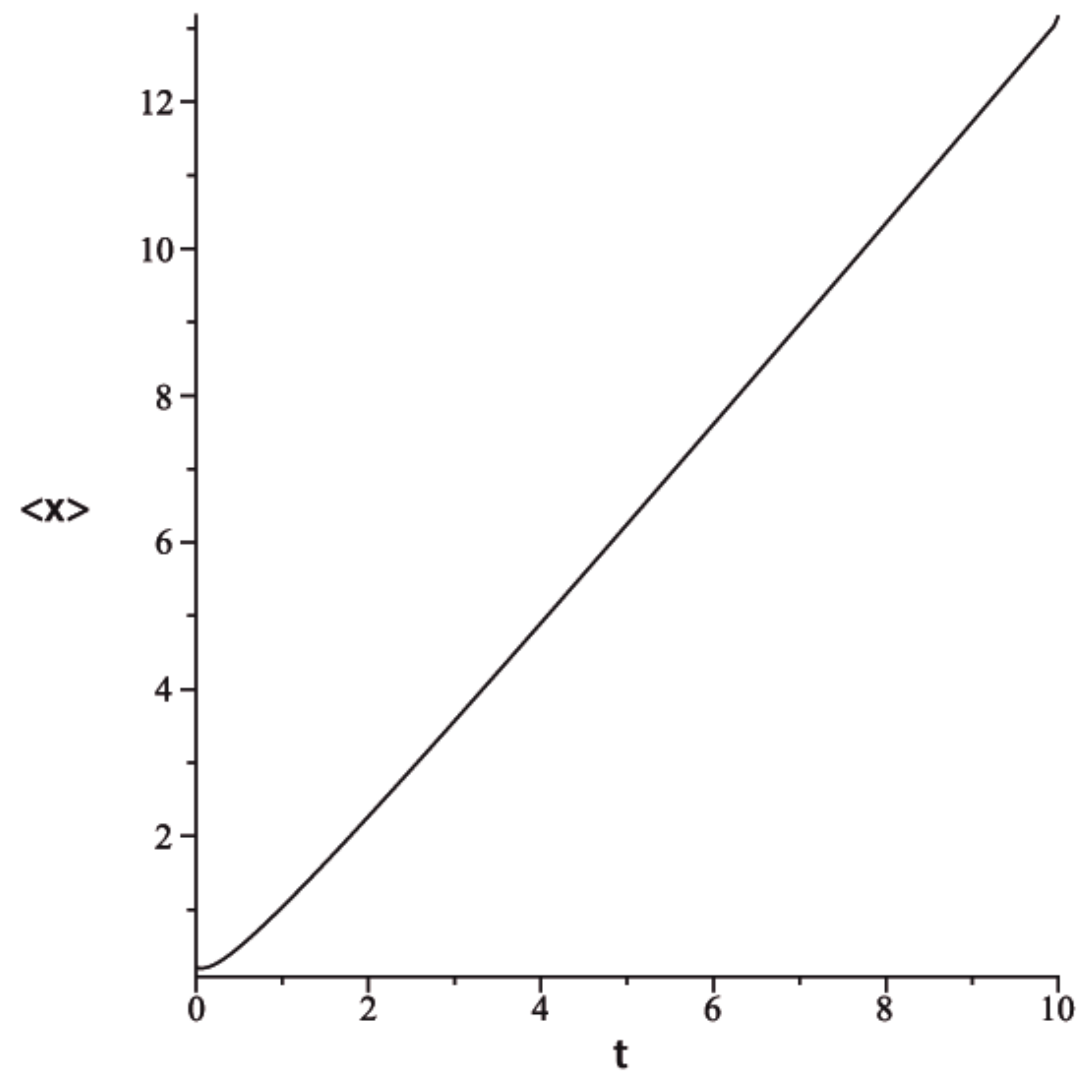}%
\caption{Temporal evolution of the expected value of the universe, constructed from the initial condition (\ref{condicao-inicial}) with energy $E_{\rm m}=3.5$.  Here we consider $dt=0.005,\, N=4500,\, x_{\rm f}=65,\, t_{\rm f}=10$. }
\label{xmedio-E3.5}
\end{center}
\end{figure}

The result shows that the universe has been expanding since its origin at a time $t_{0}$ in the past. To illustrate this, we calculated the time evolution of the expected value of the scale factor, for wave packets constructed from the initial condition (\ref{condicao-inicial}) with energy $E_{\rm m} = 3.5$, very close to the top of the potential (but smaller) barrier, as shown in Fig. \ref{xmedio-E3.5}.

We can see that $\left<x\right>(t)\neq 0$ for all value of $t$, indicating that the quantum model is free of singularities.

\subsection{The classical evolution of the scale factor of the universe}

After the Universe emerges classically to the right of the potential barrier at some point $x_2$, depending on the energy $E_m$ of the initial wave packet, the dynamics of its evolution is governed by Hamilton's equations
\begin{displaymath}
\dot{x} = \frac{\partial H}{\partial p_{x}}= -\frac{1}{12}p_{x}\, ,
\end{displaymath}
\begin{displaymath}
\dot{p_{x}} = -\frac{\partial H}{\partial x}= 
 -\frac{4\sqrt[3]{3}}{3x^{5/3}}+\frac{4}{27}{\sqrt[3]{9}}\frac{{p_{T}}_{r}}{\sqrt[3]{x^{7}}} +\frac{{p_{T}}_{d}}{3x^{2}}
\end{displaymath}

\begin{equation}
+ \frac{2\sqrt[3]{3}{p_{T}}_{cs}}{9\sqrt[3]{x^{5}}}+\frac{\sqrt[3]{9}{p_{T}}_{dw}}{9\sqrt[3]{x^{4}}}\,.
\end{equation}
They are obtained from Eq. (\ref{hnew}), and their combination gives rise to the ordinary nonlinear second order differential equation 
\begin{equation}
\frac{d^2 x}{dt^2}-\frac{\sqrt[3]{3}}{9x^{5/3}}+\frac{\sqrt[3]{9}p_{T_{r}}}{81x^{7/3}}+\frac{p_{T_{d}}}{36x^2}+\frac{\sqrt[3]{3}p_{T_{cs}}}{54x^{5/3}}+\frac{\sqrt[3]{9}p_{T_{dw}}}{108x^{4/3}} = 0\quad.
\label{eq-classica-x}
\end{equation}

By assigning the same values fo the parameter $p_T$ which have been adopted so far, the Eq. (\ref{eq-classica-x}) can be solved for the initial conditions 
\begin{equation}
x(9.95)=13.03390565\,\,\,;\,\,\, {\dot{x}(9.95)=\left.\frac{d\left<x\right>}{d\tau}\right|_{\tau = 9.95}=2.763117600}\quad,
\label{condicoes-iniciais-classicas}
\end{equation}
in which the value of $\dot{x}(9.95)$ has been obtained
numerically. The value of the rate of change of the scale factor at
instant $t_{\rm f} = 10$ will be considered equal to the value of the time
derivative of $\left<x\right>$ at that instant.

\begin{figure}[!h] 
\begin{center}
\includegraphics[height=5.5cm,width=7cm]{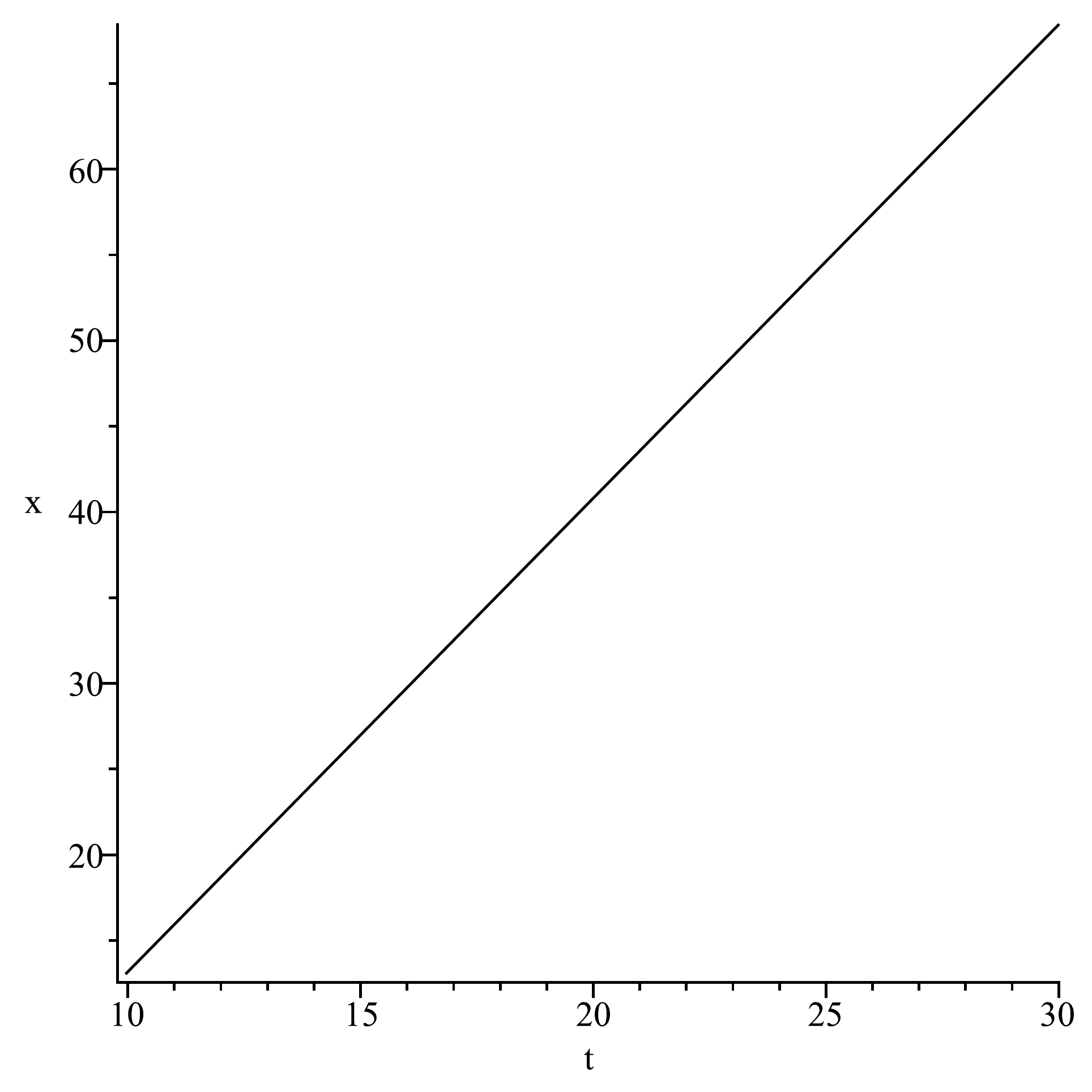}%
\caption{Classical evolution of the scale factor for the following initial conditions $x(10)=65,\,\,\,\, \dot{x}(10)=2.763117600$. }
\label{xmedioE0.1}
\end{center}
\end{figure}

In all cases numerically analyzed, we made use of the fourth order Runge-Kutta method. The result is an expanding universe. The Fig. \ref{xmedioE0.1} shows the classical evolution of the scale factor of the universe.

\section{Conclusions}
\label{CC}
In the present work we have studied the quantization of a FRW
cosmological model with positive curvature of its spatial section, the
material content of which is composed of non-interacting five perfect
fluids: radiation, dust, cosmic strings, domain walls and vacuum. 

The quantization, performed by Schultz formalism, showed that
in all cases analyzed, wave packages with finite norm could
be built which were well-defined for all space points, even
as $a\rightarrow 0$.

Employing Schutz variational method for fluids allows one to solve,
for all cases, the time problem, by choosing one
of the fluids as the ``new'' time. Except for the vacuum case, 
one obtains effective potentials that yield quantum bound states. 
For tackling those cases, Galerkin method has been applied. 
In all cases, the dynamics is described
by wave packages.

The expected value of the scale factor, but for the vacuum case,
describes primordial Universes which oscillate in a bounded interval,
with non-vanishing scale factor. For those cases, where the Universes
oscillate, we notice that the scale factor expected value oscillates, 
such that, its maxima and minima values are inside the region where the 
classical scale factor trajectory oscillates.
For the vacuum case, the expected value
of the scale factor slightly contracts (but never vanishes) at first,
and then begins an expansion process, indicating that the Universe
emerges from its quantum to its classical phase. The mechanism of such
transition is the quantum tunneling. It is striking that for the vacuum
case the tunneling probability is very close to unity, much larger than
the ones obtained in the literature for other cases \cite{monerat11, Barros}. 

We conclude that from the five fluids studied, the vacuum 
is the best candidate for the role of ``time'', being the only one that
describes a mechanism of transition from Planck's era to the classical phase.




\end{document}